\documentclass{SciPost}

\hypersetup{
    colorlinks,
    linkcolor={red!50!black},
    citecolor={blue!50!black},
    urlcolor={blue!80!black}
}

\usepackage[bitstream-charter]{mathdesign}
\usepackage{subcaption}
\DeclareSymbolFont{usualmathcal}{OMS}{cmsy}{m}{n}
\DeclareSymbolFontAlphabet{\mathcal}{usualmathcal}

\newcommand{\ptmiss}{\ensuremath{p_T^{\mathrm{miss}}}}

\fancypagestyle{SPstyle}{
\fancyhf{}
\lhead{\colorbox{scipostblue}{\bf \color{white} ~SciPost Physics Proceedings }}
\rhead{{\bf \color{scipostdeepblue} ~Submission }}

\fancyfoot[C]{\textbf{\thepage}}
}

\begin{document}

\pagestyle{SPstyle}

\begin{center}{\Large \textbf{\color{scipostdeepblue}{
Searches for Top-associated Dark Matter Production at the LHC
}}}\end{center}

\begin{center}\textbf{
Dominic Stafford\textsuperscript{1$\star$}, on behalf of the ATLAS and CMS collaborations
}\end{center}

\begin{center}
{\bf 1} Deutsches Elektronen-Synchrotron DESY, Hamburg, Germany
\\[\baselineskip]
$\star$ \href{mailto:email1}{\small dominic.william.stafford\@cern.ch}\,
\end{center}

\definecolor{palegray}{gray}{0.95}
\begin{center}
\colorbox{palegray}{
  \begin{tabular}{rr}
  \begin{minipage}{0.36\textwidth}
    \includegraphics[width=60mm,height=1.5cm]{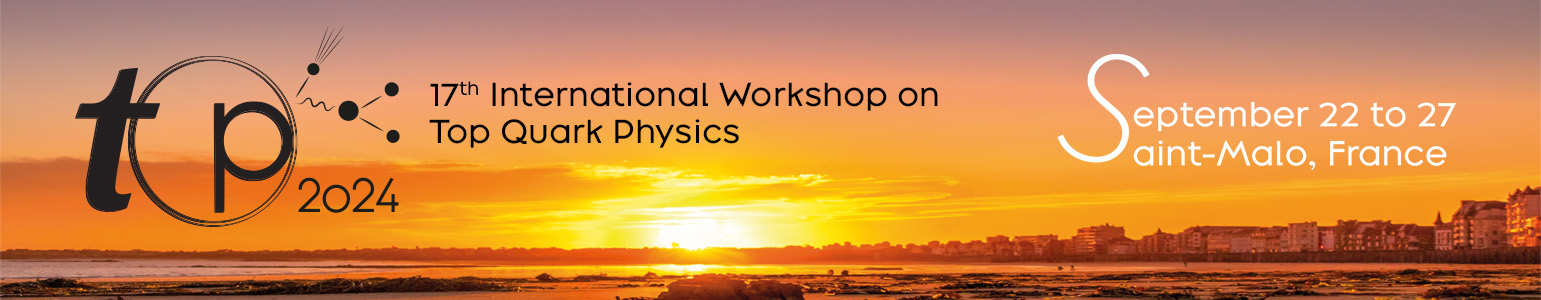}
  \end{minipage}
  &
  \begin{minipage}{0.55\textwidth}
    \begin{center} \hspace{5pt}
    {\it The 17th International Workshop on\\ Top Quark Physics (TOP2024)} \\
    {\it Saint-Malo, France, 22-27 September 2024
    }
    \doi{10.21468/SciPostPhysProc.?}\\
    \end{center}
  \end{minipage}
\end{tabular}
}
\end{center}

\section*{\color{scipostdeepblue}{Abstract}}
\textbf{\boldmath{%
Recent searches for dark matter (DM) produced in association with top quarks from the ATLAS and CMS experiments using data collected between 2015 and 2018 are presented. These comprise searches from both experiments for DM in association with a single top quark; an improved ATLAS search for DM in single lepton $t\bar{t}$ final states; an ATLAS search stop squarks decaying to a top quark, a charm quark and neutralinos, and a CMS search for DM produced in association with a pair of top quarks or a single top. These analyses feature novel machine learning and advanced background estimation techniques. No statistically significant excess is observed in any of these searches.
}}

\vspace{\baselineskip}

\noindent\textcolor{white!90!black}{%
\fbox{\parbox{0.975\linewidth}{%
\textcolor{white!40!black}{\begin{tabular}{lr}%
  \begin{minipage}{0.6\textwidth}%
    {\small Copyright attribution to authors. \newline
    This work is a submission to SciPost Phys. Proc. \newline
    License information to appear upon publication. \newline
    Publication information to appear upon publication.}
  \end{minipage} & \begin{minipage}{0.4\textwidth}
    {\small Received Date \newline Accepted Date \newline Published Date}%
  \end{minipage}
\end{tabular}}
}}
}


\vspace{10pt}
\noindent\rule{\textwidth}{1pt}
\tableofcontents
\noindent\rule{\textwidth}{1pt}
\vspace{10pt}

\section{Introduction}
\label{sec:intro}

\begin{figure}[!ht]
    \centering
    \includegraphics[width=8cm]{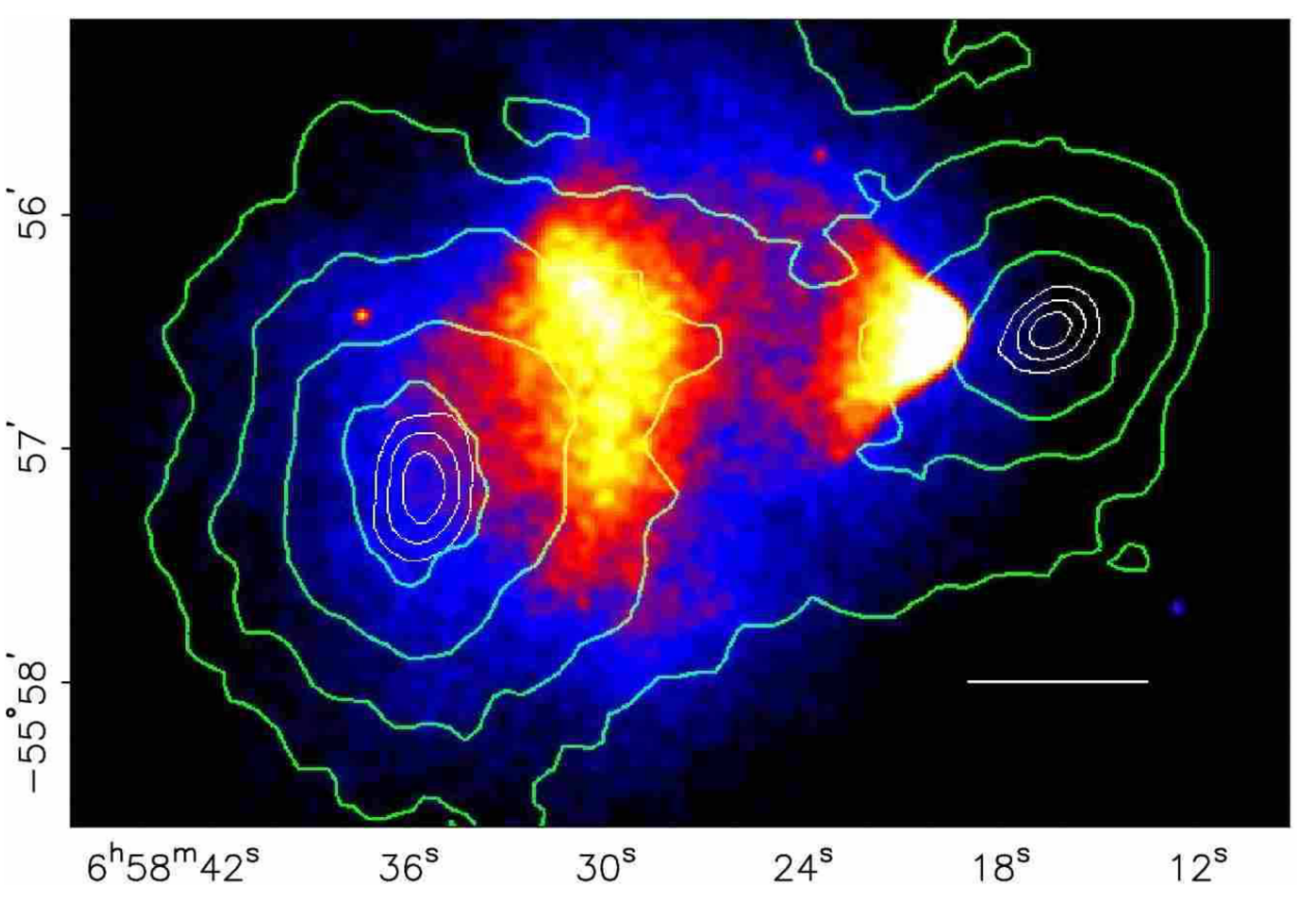}
    \caption{The bullet cluster - a pair of galaxy clusters which have collided, causing the gas clouds (colour map) to be heated and slowed by the collision, while the majority of the matter in the form of DM passes starlight through (contour, mapped by gravitational lensing of more distant objects). Source: \cite{Clowe_2006}}
    \label{fig:bullet_cluster}
\end{figure}

\begin{figure}[!ht]
    \centering
    \includegraphics[width=4.5cm]{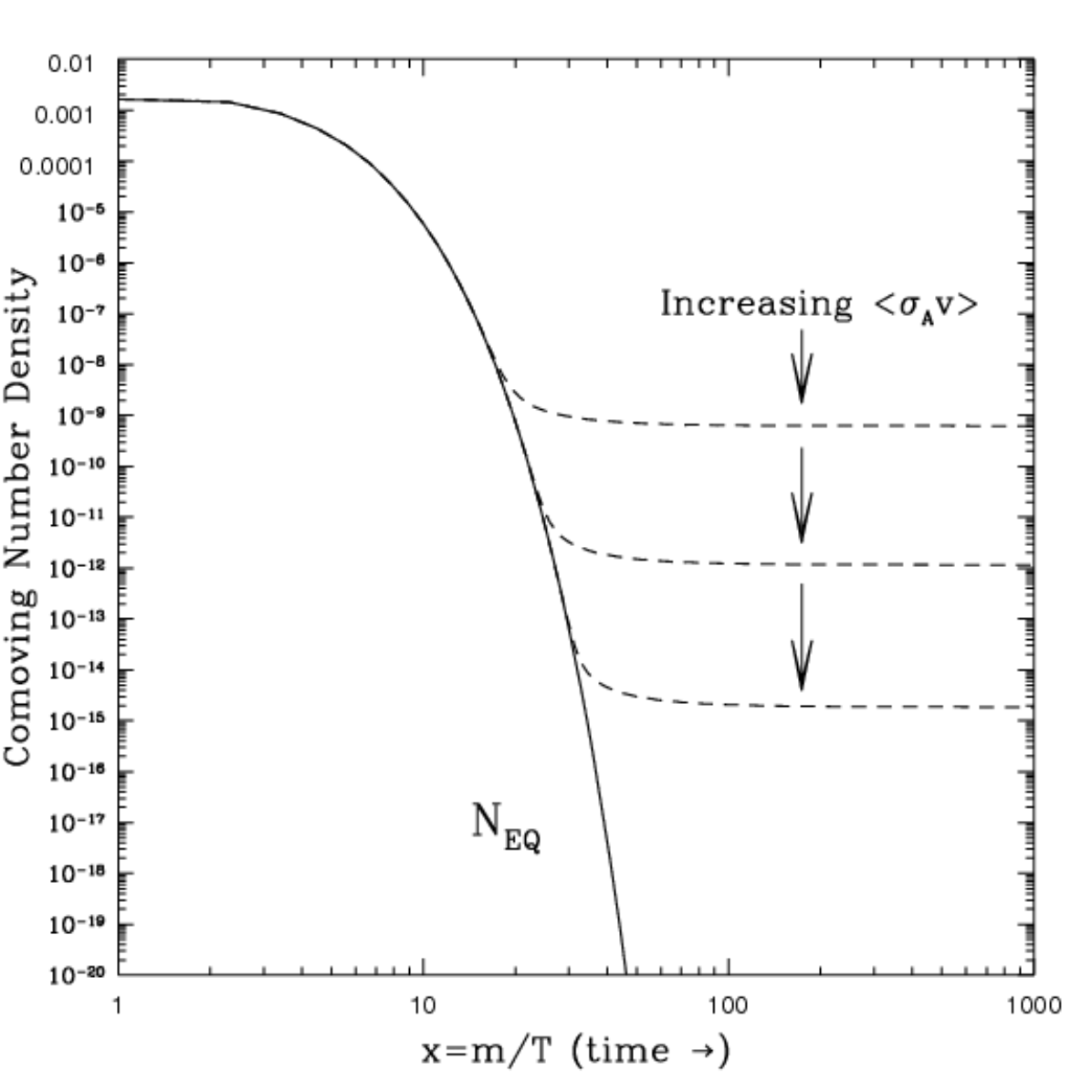}
    \caption{The density of DM as a function of time in the thermal freeze-out model. The solid line indicates the density for no freeze-out, while the dashed lines indicate three different freeze-out scenarios for different cross sections. Source: \cite{freeze_out_fig}}
    \label{fig:freeze-out}
\end{figure}

There is a large amount of evidence from astronomical and cosmological observations (for example from the bullet cluster, figure \ref{fig:bullet_cluster}) which supports the hypothesis that the majority of matter in the universe is not formed of matter described Standard Model of particle physics (SM), an is instead referred to as dark matter (DM). This dark matter is believed to account for approximately 80\% of the matter in the universe, and models of DM are usually expected to predict the density of DM in the universe today. One of the simplest models which does this is the "thermal freeze-out'' model, which proposes that DM and SM matter can interact via some high-energy interaction. In the early universe when both DM and SM particles were very dense and had a lot of energy, this interaction would often occur, and the dark and SM sectors would be in thermal equilibrium, however as the universe expanded and the energy of the particles decreased, the conversion of DM into less masive SM matter would dominate, causing the faction of the total matter in the universe in the form of DM to decrease. As the average energy of the particles decreased below that necessary for the interaction, the rate of both processes would decrease ultimately to zero, leaving some fraction of DM which could not convert to SM matter (figure \ref{fig:freeze-out}). This "relic density" would be higher for smaller DM-SM cross sections, since then the "freeze-out" will occur earlier, when less DM has converted. In particular, if one assumes the DM-SM interaction proceeds via a single s-channel mediator with order 1 couplings, this would have a mass between roughly 100 GeV and 1 TeV. This conclusion has been dubbed the "WIMP miracle", since it suggests this Dark Matter mediator may be linked to other new physics around or just above the Electroweak scale, and that it might be produced at the CERN Large Hadron Collider (LHC).


The two experiments best suited to most DM searches at the LHC are the general purpose experiments ATLAS \cite{ATLAS:2008xda} and CMS \cite{CMS:2008xjf}. These experiments use two methods to search for these types of mediators: one can search for the mediator decaying back to SM particles, or one can search for events where the mediator decays to dark matter. The former is more sensitive to cases where the mediator couplings to the SM are similar to or larger than those to DM, and many such models have been excluded by resonance searches, which will not be covered here. The latter is more challenging since the DM thus produced cannot be directly detected, however if the dark matter is produced in association with some SM particles, one can infer the presence of DM from the fact it will tend to recoil against these SM particles. Since the initial state particles have almost no momentum in the plane transverse to the colliding beams, the negative vector sum of the transverse momenta of all visible particles, referred to as "missing transverse momentum" or \ptmiss{}, can be assigned to invisible particles, including DM.

Since the new DM mediator predicted by the thermal freeze-out model has a mass around the electroweak scale, many DM models also relate to other new physics at this scale, for instance Higgs sector extensions. This often leads to models in which DM is produced in association with top quarks, the heaviest particles in the SM. Examples include new bosons with enhanced couplings to top quarks, and stop squarks, supersymmetric particles which decay to top quarks and a neutral particle which can be a DM candidate.

\section{Mono-top searches}
\label{sec:monotop}

\begin{figure}[!ht]
    \centering
    \includegraphics[width=6cm]{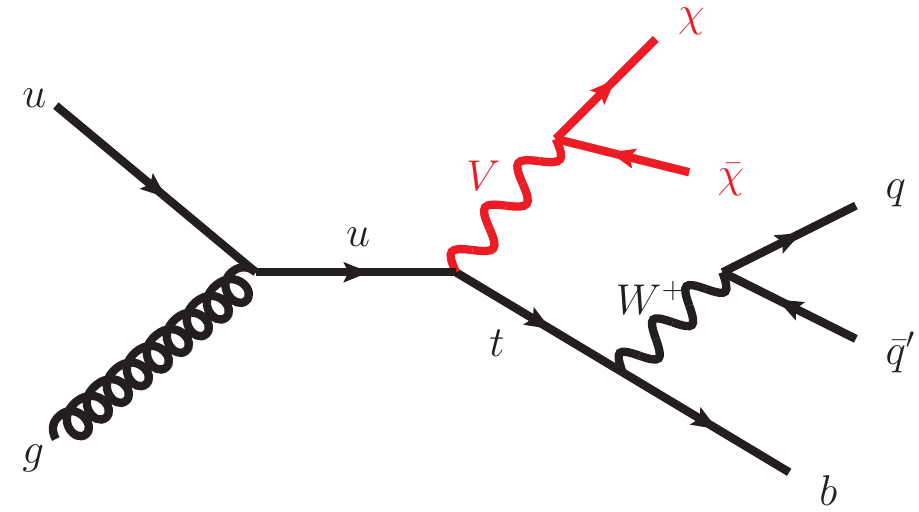}
    \caption{Production of DM in association with a single top quark via emission of a FCNC mediator. Source: \cite{ATLAS:2024xne}}
    \label{fig:monotop_tch_diag}
\end{figure}

Some dark matter models include flavour-changing neutral current (FCNC) mediators, which can change the flavour of the quark emitting them, and then decay to DM, which can give a distinctive signature of a single boosted top quark recoiling against \ptmiss{}, as shown in figure \ref{fig:monotop_tch_diag}. Previous studies have found the channel in which the top quark decays hadronically and can be reconstructed as a single large radius jet is most sensitive to this signal.

\begin{figure}[!ht]
    \centering
    \includegraphics[width=8cm]{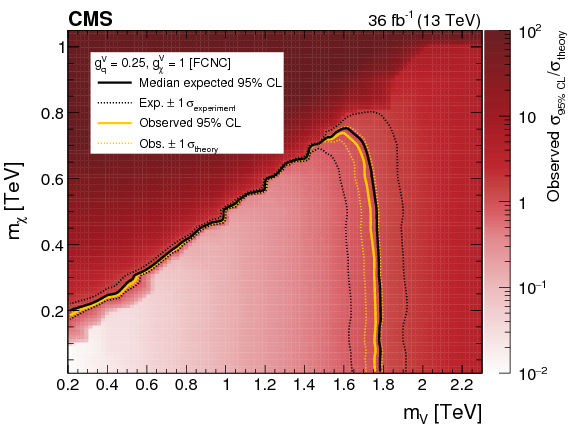}
    \caption{Previous limits on t-channel production of DM in association with a top quark for a vector FCNC mediator from \cite{CMS:2018gbj}}
    \label{fig:CMS_2016_monotop_lims}
\end{figure}

The main challenge for these analyses are the numerous SM backgrounds, which are not always well modelled in Monte Carlo Simulation, and hence require dedicated background estimation techniques. Until this year, the most sensitive searches were those from ATLAS \cite{ATLAS:2018cjd} and CMS \cite{CMS:2018gbj} with the data collected in 2016, which excluded mediator masses up to about 1.8 TeV (figure \ref{fig:CMS_2016_monotop_lims}). In the past year both experiments have published searches using the full Run 2 dataset (data collected at 13 TeV between 2015 and 2018) \cite{ATLAS:2024xne, CMS:2024tkt}, which are presented here.

\subsection{Search strategies}
\label{sec:monotop_strat}

Both new searches select events with high \ptmiss{} ($>250$ GeV for ATLAS, $>350$ GeV for CMS), a single large-radius jet clustered with the anti-kT algorithm (radius 1.0 and $p_{T}>350$ GeV for ATLAS and radius 1.5 and $p_{T}>250$ GeV for CMS), and veto events with isolated leptons. Similarly, both analyses use Deep Neural Networks (DNNs) to select large-radius jets coming from top quarks - ATLAS requires the large-radius jet passes the medium working point of a DNN trained on jet kinematics and substructure variables, while CMS uses particleNet, a graph neural network trained on all of the particles within a jet, to split the signal region (SR) into a signal enriched top-pass category, and a less enriched top-fail category.

\begin{figure}[htbp]
\begin{minipage}[b]{.29\textwidth}
\begin{center}
\includegraphics[height=5cm]{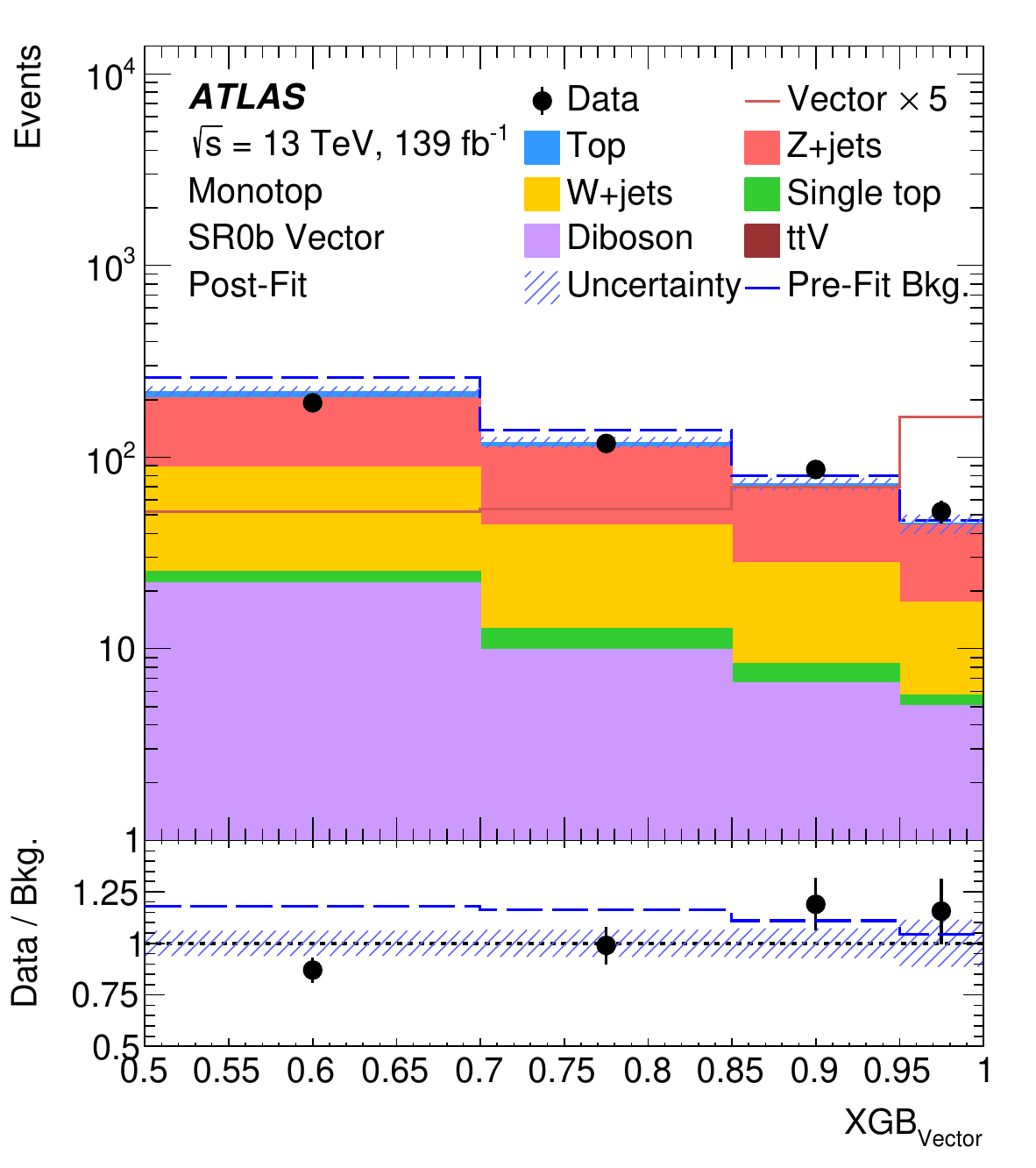}
\end{center}
\end{minipage} \hfill
\begin{minipage}[b]{.29\textwidth}
\begin{center}
\includegraphics[height=5cm]{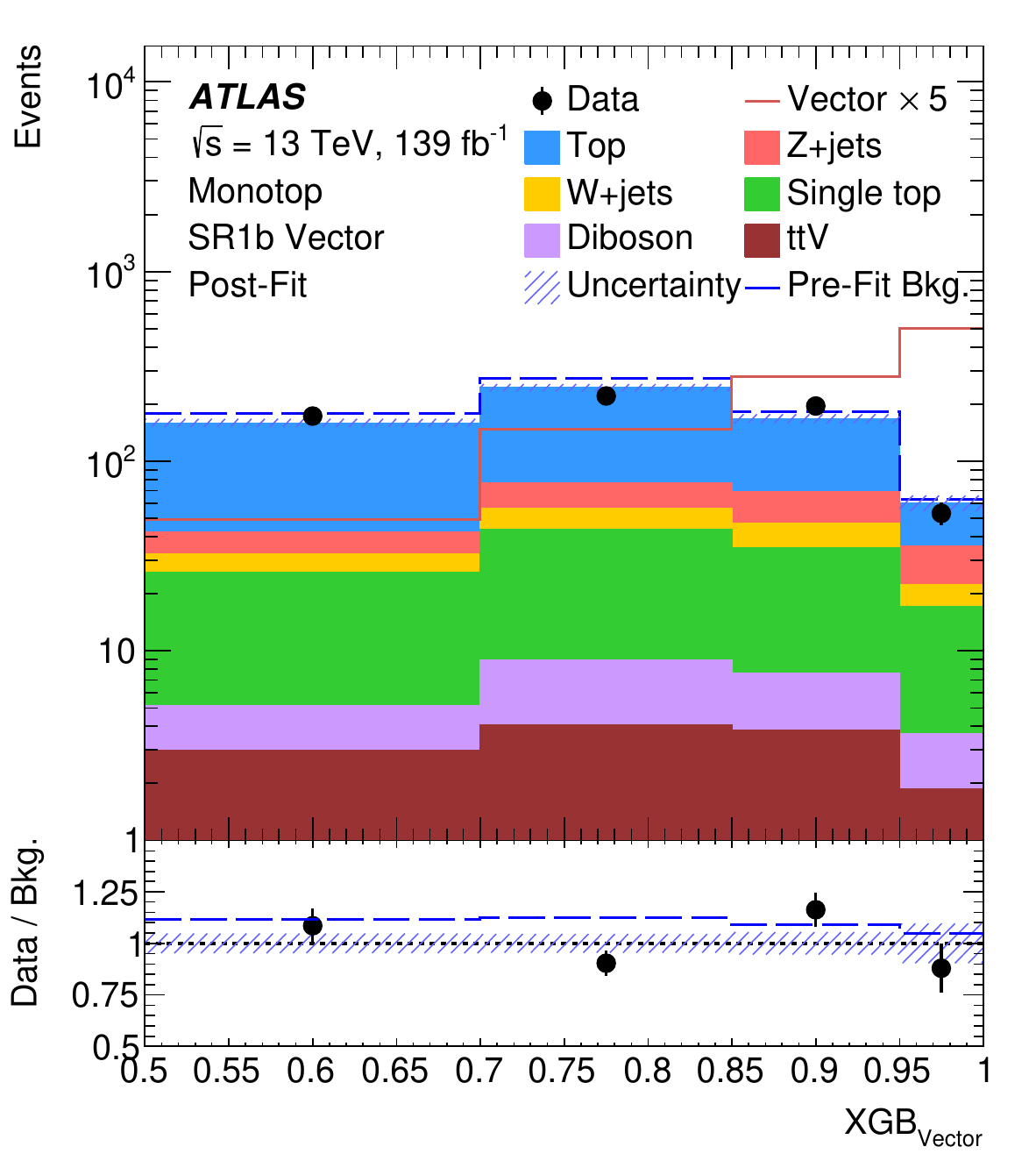}
\end{center}
\end{minipage} \hfill
\begin{minipage}[b]{.39\textwidth}
\begin{center}
\includegraphics[height=5.4cm]{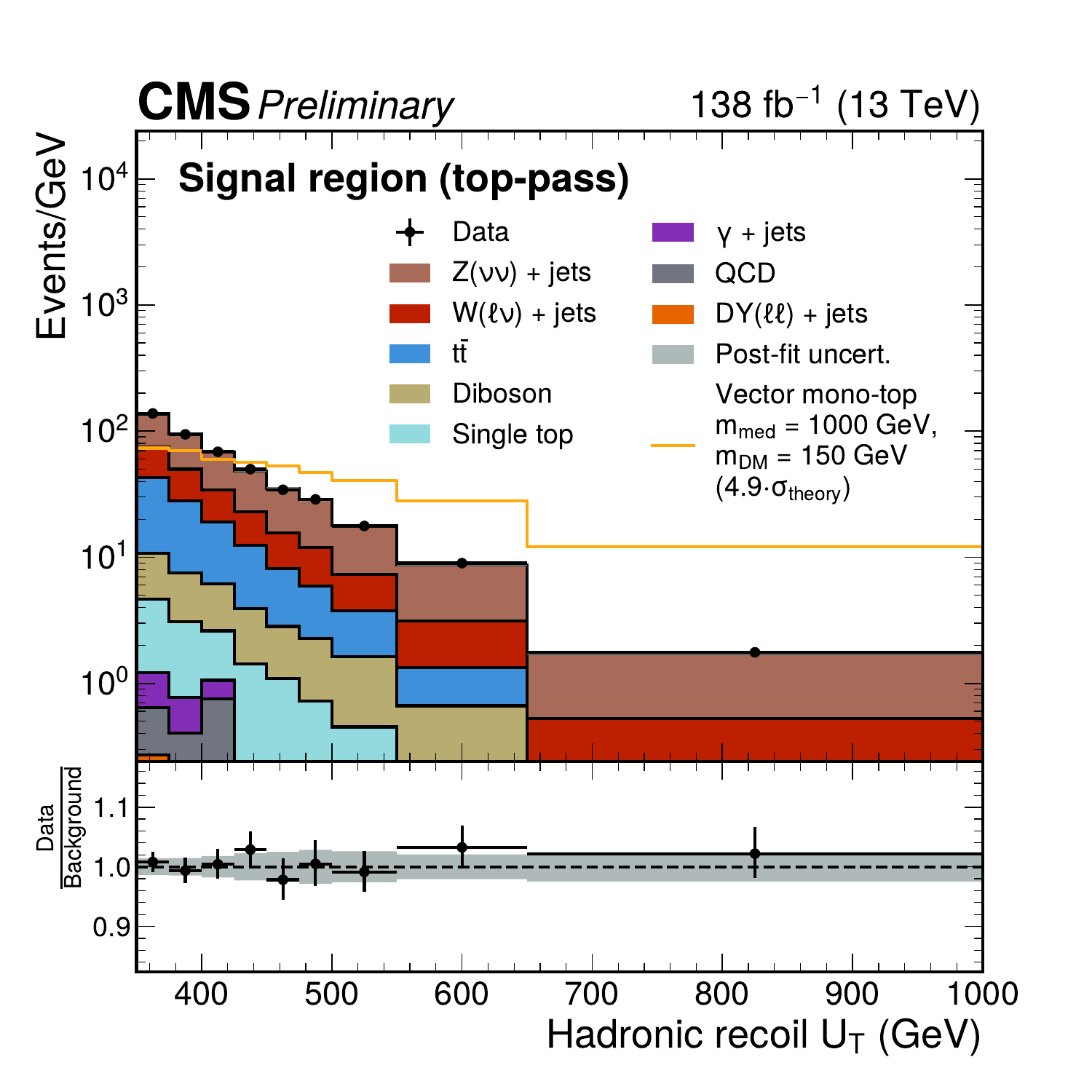}
\end{center}
\end{minipage}
\caption{Signal regions for the ATLAS 0b (left) and 1b (centre) categories, and for the CMS signal-enriched top-pass category (right)}
\label{fig:monotop_SRs}
\end{figure}

Where the analyses differ is the fitting strategy- here CMS fit on directly on the \ptmiss{} in the top-pass and -fail categories, but ATLAS employ a more complex strategy. Firstly, the SR is split according to the number of b-tagged small radius jets (which may be inside the large-radius jet), with categories for 0 and 1 b-tagged jets. Then a boosted decision tree (BDT), called XGBoost, was trained on the kinematics of the \ptmiss{} and large-radius jet, as well as those of any additional small-radius jets in the event. The SR was required to have a BDT score > 0.5, and the fit was performed on this variable. These distributions are shown in figure \ref{fig:monotop_SRs}.

\subsection{Background estimation}
\label{sec:monotop_bkgd}

A second factor in the optimisation of these analyses is the design of the control regions used to ensure good modelling of the backgrounds. The largest backgrounds are $Z\to\nu\bar{\nu}$ production, where initial state radiation (ISR) gives a large radius jet and the neutrinos are the source of the \ptmiss{} (and to a lesser extent $W\to l\nu$ production, where the lepton is not identified), and single lepton $t\bar{t}$ decays, where the large radius jet comes from the hadronically decaying top quark, the lepton is not identified, and the neutrino from the leptonic top decay is the source of \ptmiss{}.

\begin{figure}[!ht]
    \centering
  \begin{subfigure}[t]{.48\textwidth}
    \centering
    \includegraphics[width=1\textwidth]{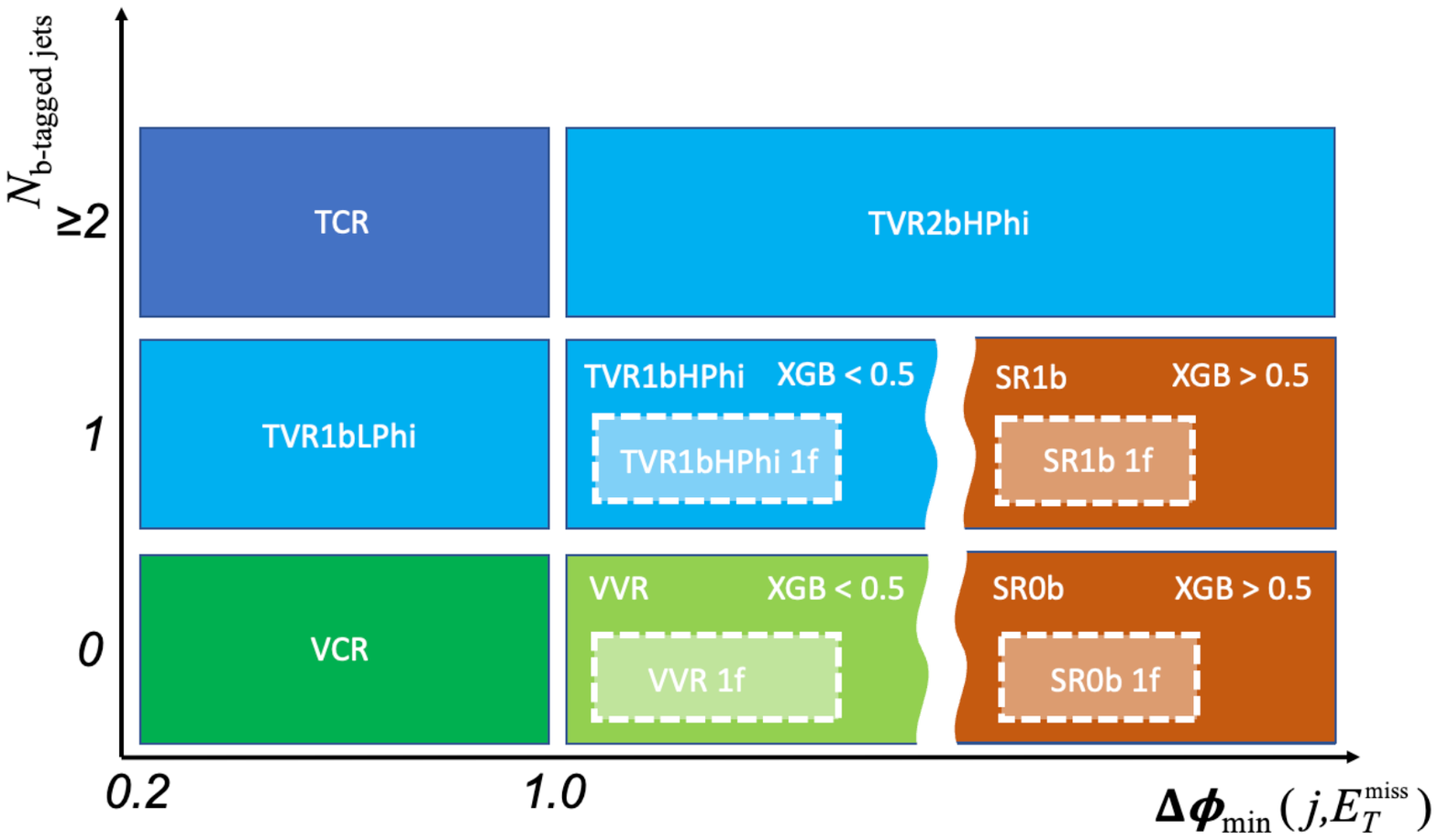}
    \caption{The control, validation and signal regions for the ATLAS analysis}
    \label{fig:ATLAS_monotop_CR_defs}
\end{subfigure}\hfill
  \begin{subfigure}[t]{.48\textwidth}
    \centering
    \includegraphics[width=1\textwidth]{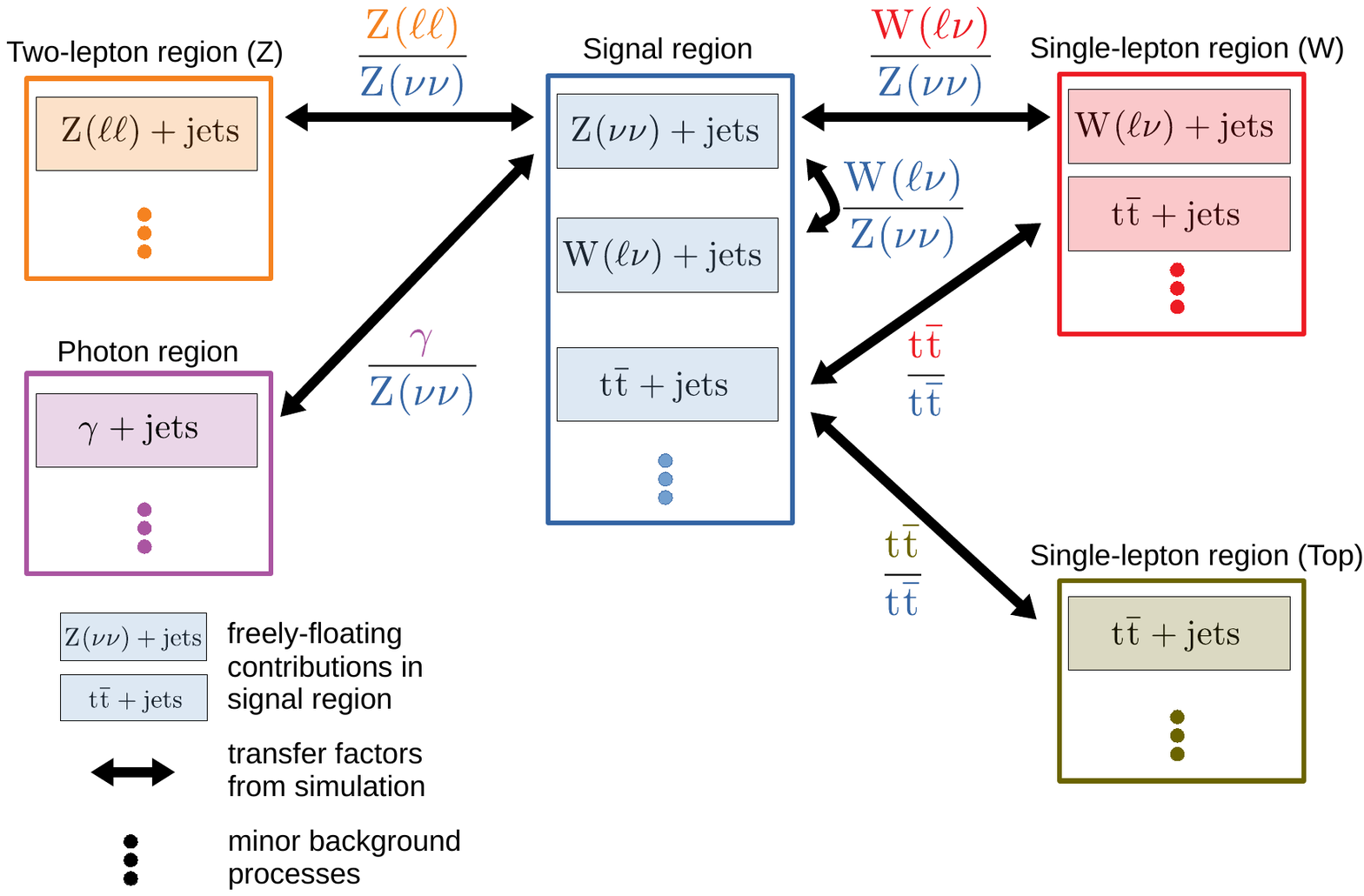}
    \caption{The signal and control regions in the CMS analysis, including all transfer factors used to link the rates of different processes in the fit}
    \label{fig:CMS_monotop_CR_defs}
\end{subfigure}\hfill
\end{figure}

ATLAS define control regions using a selection on the angle between the large-radius jet and \ptmiss{}, $0.2<\Delta\phi(j,\ptmiss)<1.0$ (while the signal region has a cut of $\Delta\phi(j,\ptmiss)>1.0$ since in the signal the top quark recoils against the invisible DM, with little other radiation). The region with exactly 0 b-tagged jets is used to estimate the rate of $Z\to\nu\bar{\nu}$ production, while the the region with 2 or more b jets is used to estimate the rate of $t\bar{t}$ production. A series of validation regions (VRs) are then used to ensure good modelling of this background, specifically the $0.2<\Delta\phi(j,\ptmiss)<1.0$, 1 b-tagged jet region (to validate the rate of $t\bar{t}$ when one b-quark is not tagged); the $\Delta\phi(j,\ptmiss)>1.0$, 2 b-tagged jet ($t\bar{t}$ in a signal-like topology), and finally the $\Delta\phi(j,\ptmiss)>1.0$, 0 and 1 b-tagged jet, BDT score < 0.5 regions, which differ from the SRs only in the BDT cut. These regions are summarised in figure \ref{fig:ATLAS_monotop_CR_defs}.

CMS instead define control regions by requiring the presence of leptons in addition to the large-radius jet; for Z boson production this amounts to estimating the rate of $Z\to\nu\bar{\nu}$ from $Z\to l^{+}l^{-}$ in a region with 2 leptons and $\ptmiss<120$ GeV, whilst for single lepton $t\bar{t}$ and $W\to l\nu$ production it corresponds to estimating the rate where the lepton is not identified from regions with 1 lepton, $\ptmiss>150$ GeV and $\geq 1$ and 0 b-tagged jets, respectively. Furthermore, the rate is estimated separately in each bin of the hadronic recoil (the negative vector sum of only the hadronic components of the event), which corresponds to \ptmiss{} in the SR, allowing fine-grained estimation of the backgrounds as a function of the fit variable. Due to the similarity of the processes, the rates of Z and W boson production are linked, and an additional CR targeting photon + jet production was added to give a further handle on the rate of these boson + jet processes. The rates of the processes were allowed to float in the final fit, being constrained by the CR data - the different control regions, and the transfer factors linking them, are shown in figure \ref{fig:CMS_monotop_CR_defs}.

\subsection{Results}
\label{sec:monotop_results}

\begin{figure}[htbp]
\begin{minipage}[b]{.29\textwidth}
\begin{center}
\includegraphics[height=5cm]{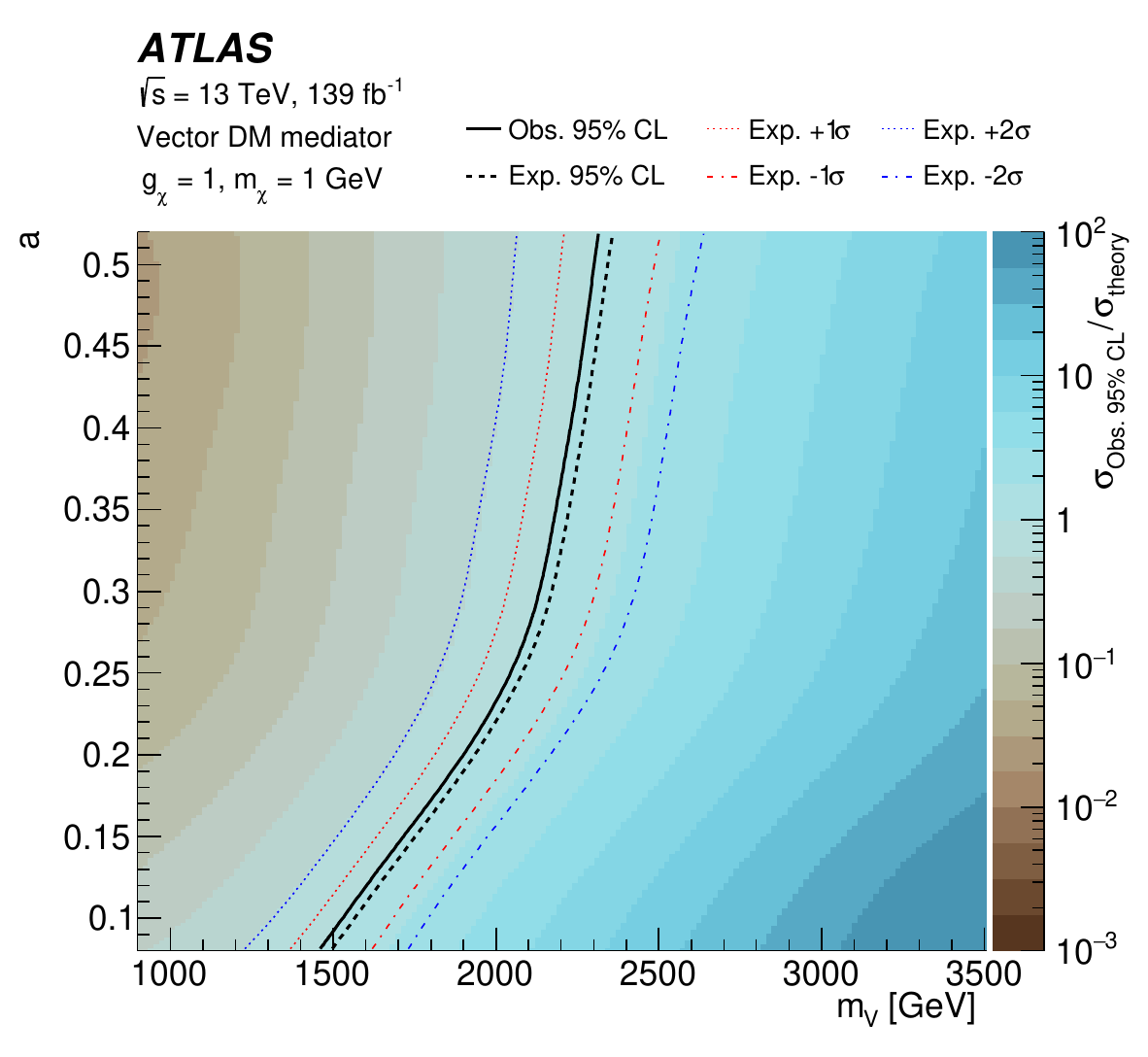}
\end{center}
\end{minipage} \hfill
\begin{minipage}[b]{.29\textwidth}
\begin{center}
\includegraphics[height=5cm]{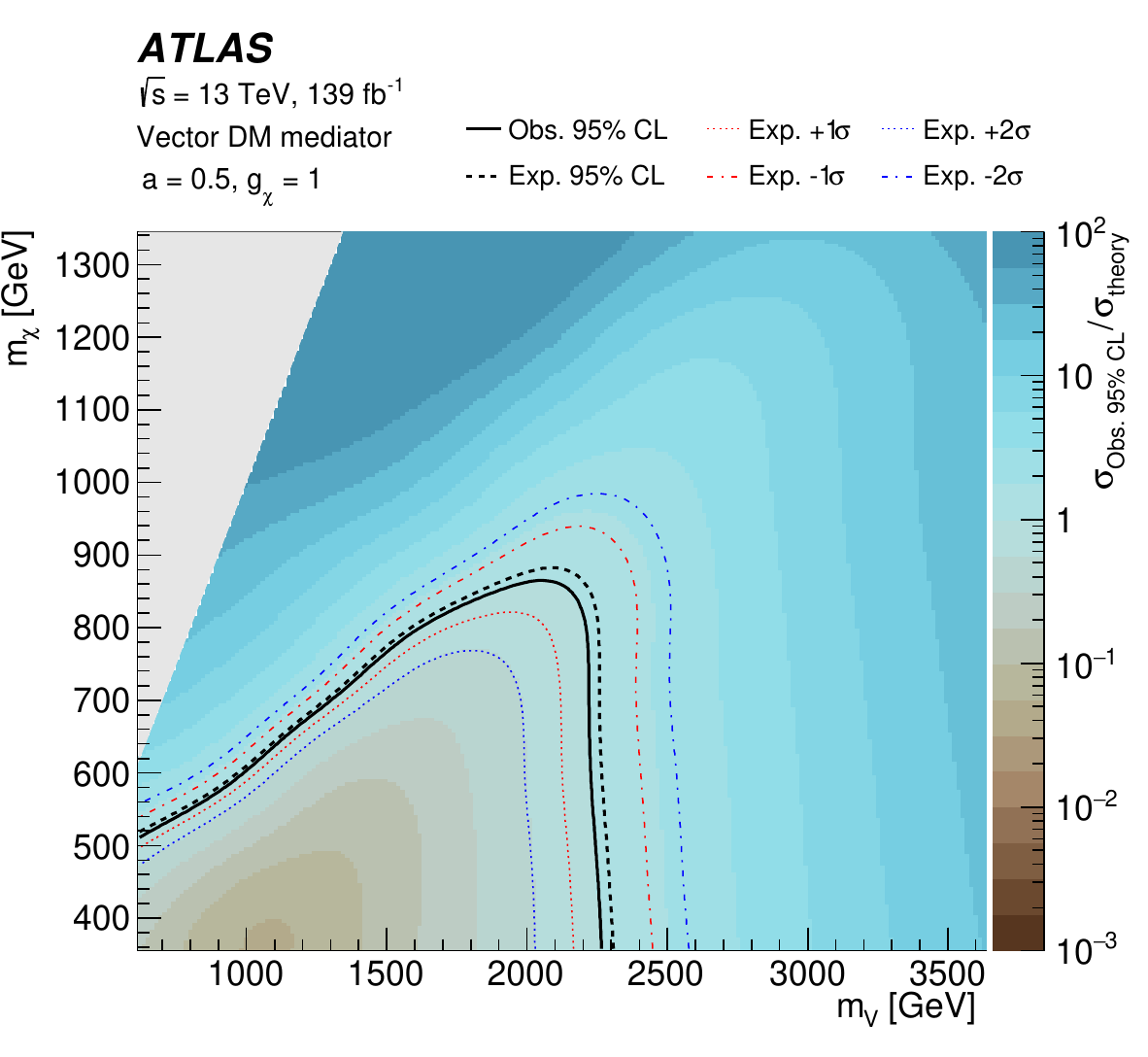}
\end{center}
\end{minipage} \hfill
\begin{minipage}[b]{.39\textwidth}
\begin{center}
\includegraphics[height=5cm]{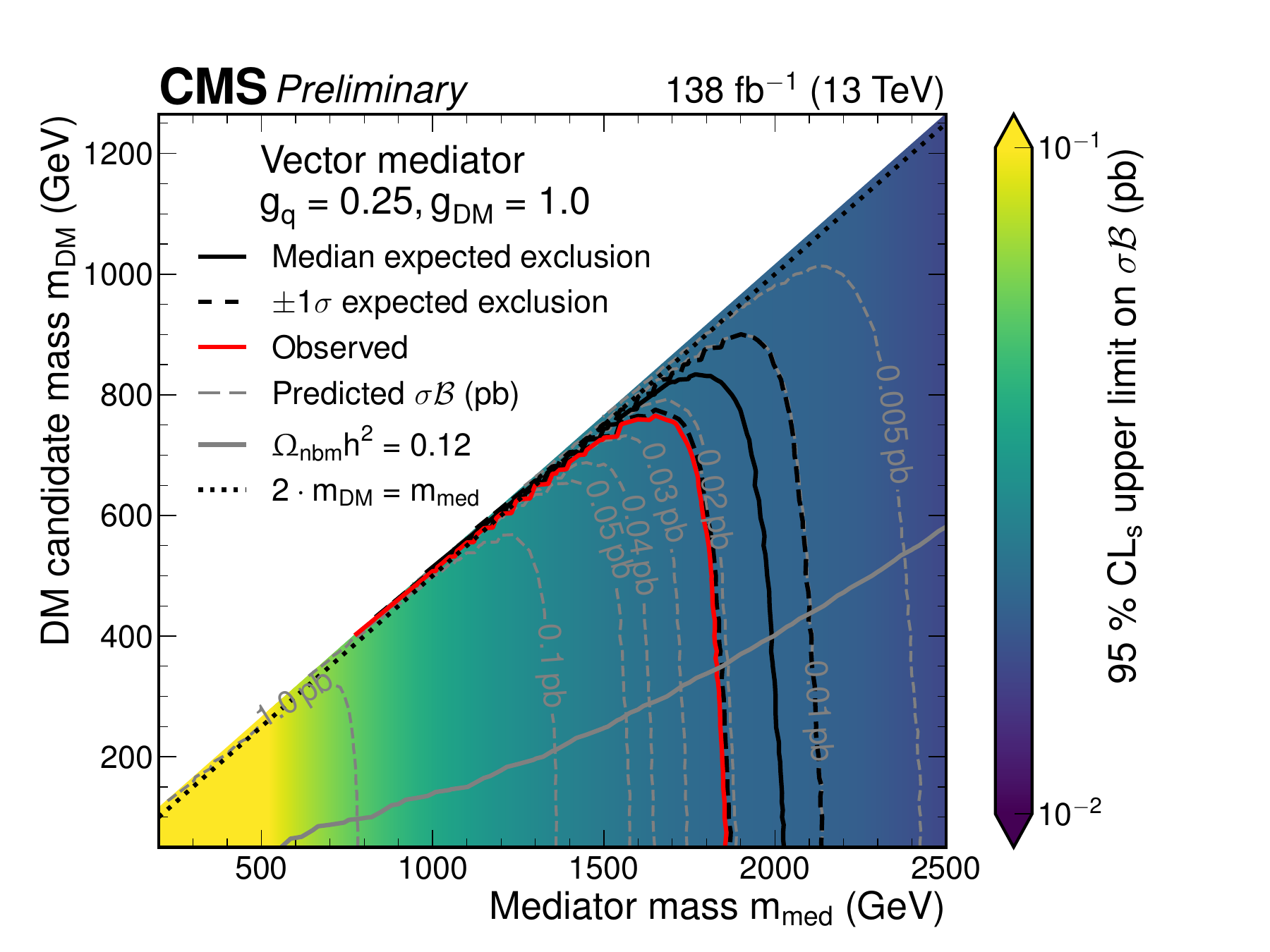}
\end{center}
\end{minipage}
\caption{Limits on different combinations of model parameters: left, the ATLAS limits on the mass of the mediator, $m_{V}$, and the coupling to the SM, $a$, for DM mass, $m_{\chi}=1$ GeV; centre, the ATLAS limits on $m_{V}$ and $m_{\chi}=1$ for $a=0.5$; right, the CMS limits on mediator mass, $m_{\mathrm{med}}$, and DM mass, $m_{\mathrm{DM}}$, for a coupling to the SM of 0.25. All plots have a coupling to DM of 1.0}
\label{fig:monotop_lims}
\end{figure}

The ATLAS search did not observe any signal-like excess for this model, and the CMS search observed only a 1-sigma excess, so both analyses set lower limits on possible combinations of DM mass, mediator mass and couplings. These are shown in figure \ref{fig:monotop_lims}; CMS performed a scan over mediator mass and DM mass for a coupling to quarks of 0.25 and to DM of 1.0, while ATLAS performed the same scan for a coupling to quarks of 0.5 and to DM of 1.0, as well as a scan over quark coupling and mediator mass for a DM mass of 1 GeV. Comparing these one can see that the for a coupling to DM of 0.25 and a DM mass of 1 GeV, the ATLAS analysis is expected to exclude mediator masses up to 80 GeV higher than the CMS search, and in the observed results the difference is even larger due to the excess in CMS data weakening the observed limits. However from the scans of DM mass against mediator mass one can see that for higher mediator masses CMS is able to exclude more closely to the kinematic limit of $2m_{\mathrm{DM}}=m_{\mathrm{med}}$, where the mediator starts to go off-shell. This may be due to the larger jet radius used by CMS giving access to less boosted topologies.

\begin{figure}[!ht]
\begin{minipage}[b]{.48\textwidth}
\begin{center}
\includegraphics[width=5cm]{Figs/monotop_tch_diag.pdf}
\end{center}
\end{minipage} \hfill
\begin{minipage}[b]{.48\textwidth}
\begin{center}
\includegraphics[width=5cm]{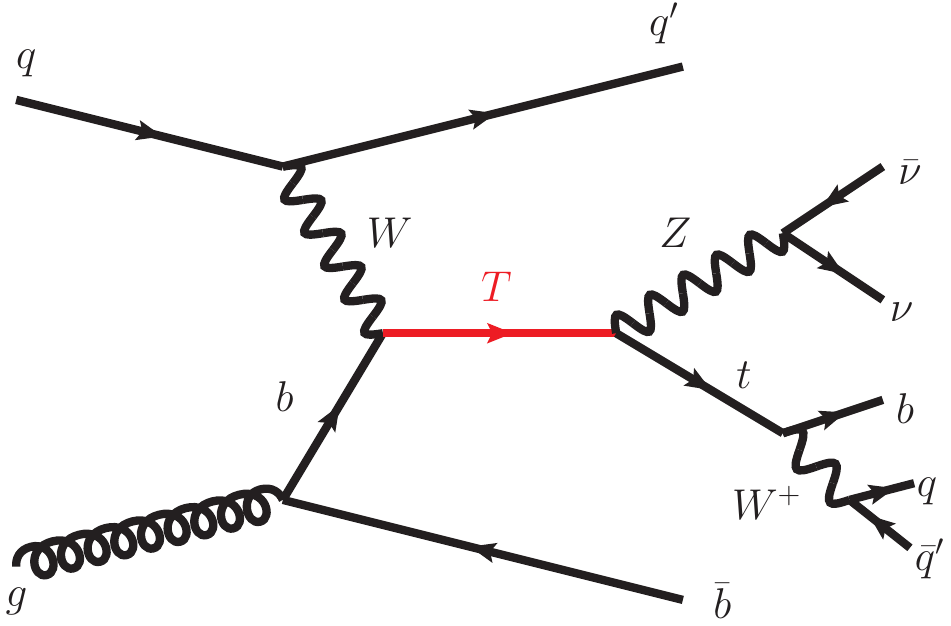}
\end{center}
\end{minipage} 
\caption{Production of DM in association with a single top quark via a resonant mediator (left), and a VLQ, T, decaying to a top quark and an invisibly decaying Z boson (right).}
\label{fig:alt_monotop_diags}
\end{figure}

\begin{figure}[!ht]
\begin{minipage}[b]{.48\textwidth}
\begin{center}
\includegraphics[width=6cm]{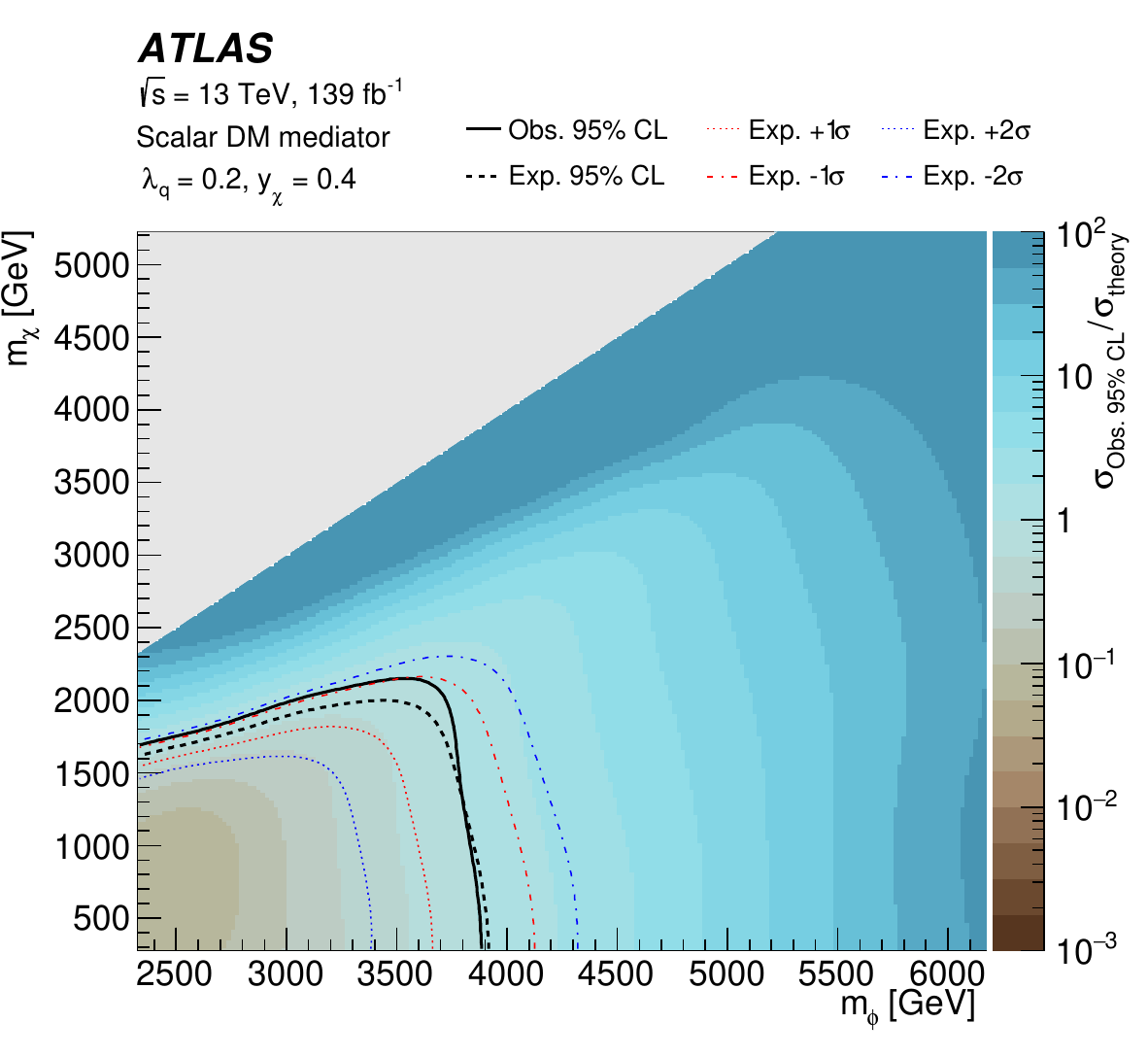}
\end{center}
\end{minipage} \hfill
\begin{minipage}[b]{.48\textwidth}
\begin{center}
\includegraphics[width=6cm]{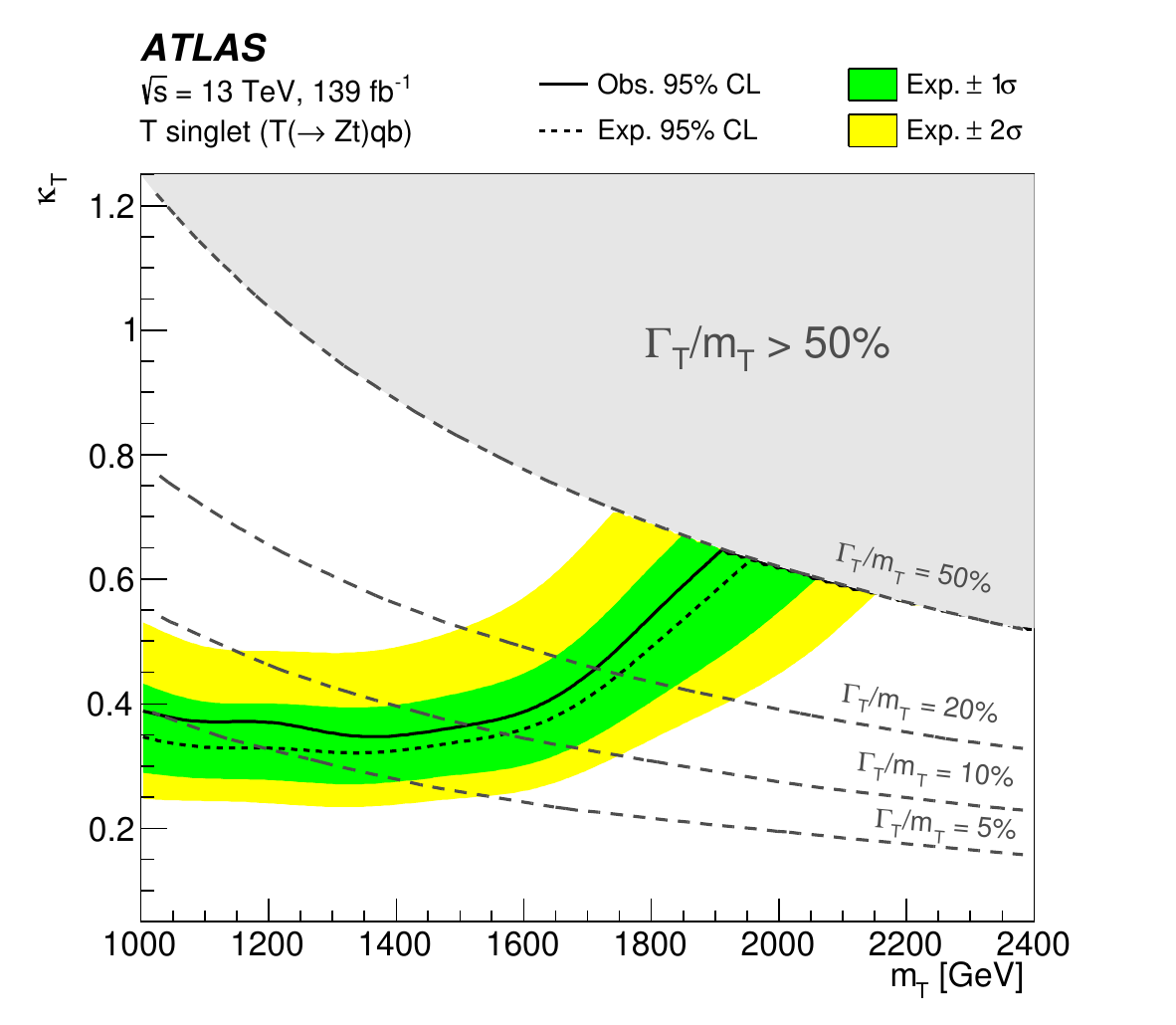}
\end{center}
\end{minipage} 
\caption{Limits on mono-top+DM as a function of the mass of the mediator, $m_{V}$, and the DM mass, $m_{\chi}$, (left) and on VLQ production as a function of the VLQ mass, $m_{t}$, and the universal coupling, $\kappa_{T}$}
\label{fig:alt_monotop_lims}
\end{figure}

In addition to this model, the ATLAS search was optimised for two further signals: firstly a resonant mediator which decays to a top quark as well as DM candidate, and secondly a vector-like quark (VLQ) which decays to a top quark and a Z boson (which then decays to $\nu\bar{\nu}$). The Feynman diagrams for these processes are shown in \ref{fig:alt_monotop_diags}. Separate BDTs were trained for each signal using different variables, and the VLQ search featured an additional categorisation in the number of forward jets, since these are often present in the diagrams which produce this signal. Again no significant excess is observed for these models, and so limits are set, as shown in figure \ref{fig:alt_monotop_lims}.

\section{Dark Matter in $t\bar{t}$ Final States}
\label{sec:ttDM}

\begin{figure}[!ht]
    \centering
  \begin{subfigure}[t]{.48\textwidth}
    \centering
    \includegraphics[width=5cm]{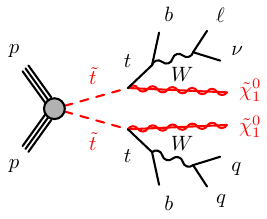}
    \caption{Production of a pair of stop squarks decaying to top quarks and neutralinos.}
    \label{fig:stop_pair_diag}
\end{subfigure}\hfill
  \begin{subfigure}[t]{.48\textwidth}
    \centering
    \includegraphics[width=5cm]{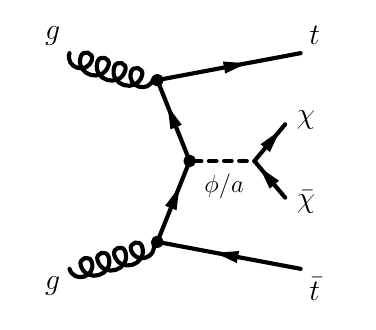}
    \caption{Production of DM in association with $t\bar{t}$ via a spin-0 mediator.}
    \label{fig:ttDM_simp_diag}
\end{subfigure}\hfill
\end{figure}

Another important signature for LHC DM searches is the production of DM in association with a $t\bar{t}$ pair. Historically one of the most commonly discussed models producing this signature is R-parity conserving supersymmetry, where a pair of stop squarks are produced, which each decay to a top quark and a neutralino, the lightest supersymmetric particle (LSP), which in these models is a stable DM candidate, as shown in figure \ref{fig:stop_pair_diag}. The phase space for these models is parameterised by mass of the stop, $m_{\widetilde{t}}$, and the mass of the neutralino, $m_{\widetilde{\chi_{1}^{0}}}$ - if $m_{\widetilde{t}}$ is much larger than $m_{\widetilde{\chi_{1}^{0}}}$ this will give signatures in which the top quarks are very boosted and the events contain a large amount of \ptmiss{}, however as $m_{\widetilde{t}}$ approaches $m_{\widetilde{\chi_{1}^{0}}}+m_{t}$ (the "compressed'' region) the tops become less boosted and the additional \ptmiss{} decreases, making these events more challenging to distinguish from SM $t\bar{t}$ production.

A set of newer models predicting a similar signature are extended Higgs sectors, which predict new scalar and pseudoscalar mediators which have Yukawa couplings, and hence will couple most strongly to the top quark out of all the SM particles, since it is the most massive SM particle. This will also give rise to $t\bar{t}$+DM signatures, as show in figure \ref{fig:ttDM_simp_diag}. These models tend to have lower mediator masses and hence less boosted topologies than other models discussed here. There a large number of such possible Higgs extensions, but to keep the results general searches normally focus on simplified models with only a single additional scalar or pseudoscalar boson.

\begin{figure}[htbp]
\begin{minipage}[b]{.48\textwidth}
\begin{center}
\includegraphics[height=6cm]{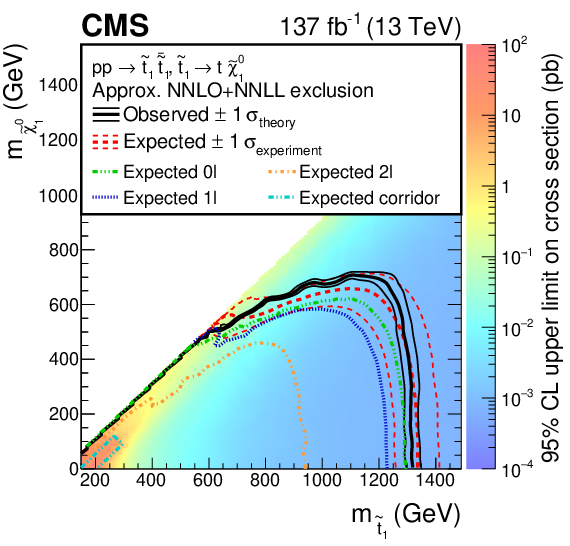}
\end{center}
\end{minipage} \hfill
\begin{minipage}[b]{.48\textwidth}
\begin{center}
\includegraphics[height=6cm]{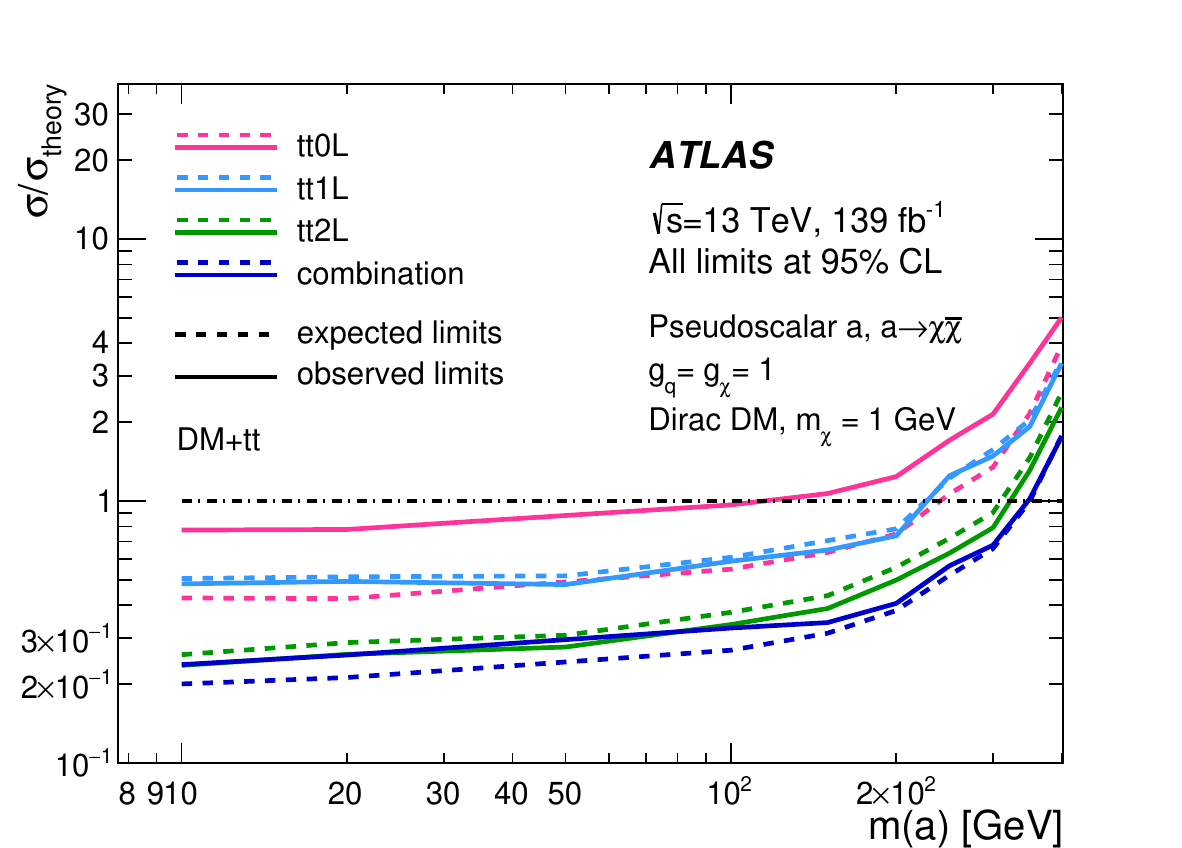}
\end{center}
\end{minipage} 
\caption{Previous limits on DM produced in association with top quarks. Left: limits on stop quark production from \cite{CMS:2021eha}, right: limits on the simplified $t\bar{t}$+DM model with a pseudoscalar mediator from \cite{ATLAS:2022ygn}.}
\label{fig:old_ttDM_lims}
\end{figure}

Both ATLAS and CMS have already performed searches for both of these models using the full run 2 datasets; the CMS limits on stop production and the ATLAS limits on $t\bar{t}$+DM via a pseudoscalar mediator are shown in figure \ref{fig:old_ttDM_lims}. However there was still room for sensitivity improvements in this dataset, and so both collaborations have released new results featuring novel analysis strategies in the past year.

\subsection{Interlude: A measurement of Standard Model $t\bar{t}$ neutrino kinematics}

\begin{figure}[!ht]
    \centering
    \includegraphics[width=8cm]{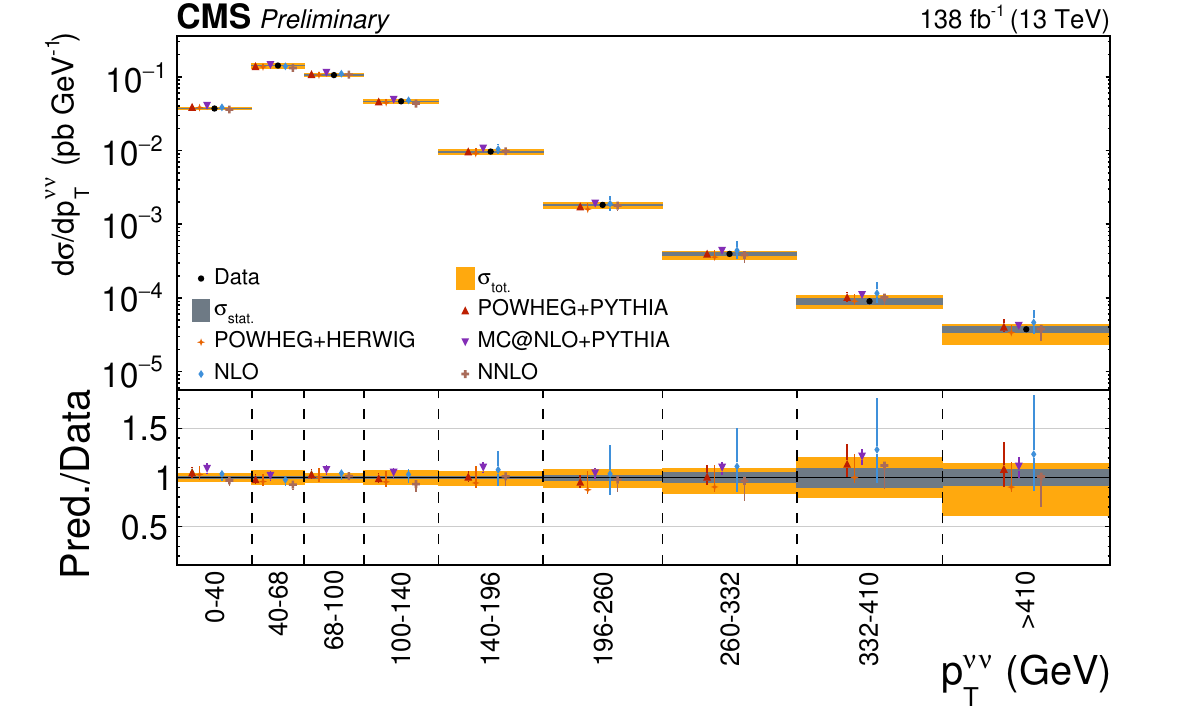}
    \caption{The dineutrino transverse momentum, $p_{T}^{\nu\nu}$, distribution for $t\bar{t}$, corrected for detector effects with the other backgrounds subtracted.}
    \label{fig:CMS_pTnunu_unfolded}
\end{figure}

The first analysis of interest is not a BSM search, but a measurement of a dileptonic $t\bar{t}$ production \cite{CMS:2024tgh}. This process contains a large amount of \ptmiss{} due to the two neutrinos in the final state, which makes it a challenging background for not only searches for dileptonic $t\bar{t}$ +DM, but also single lepton $t\bar{t}$+DM, as it can enter this channel if one lepton is not reconstructed, and then has more \ptmiss{} than would be expected from single lepton $t\bar{t}$ production. Due to the importance of this background, CMS measured the rate of this background differentially as a function of the dineutrino kinematics, specifically the transverse momentum of the dineutrino pair and the angle between this pair and the nearest lepton. These results were presented in more detail in \cite{tt_BSM_PS} - good agreement was seen with all generators, as can be seen in figure \ref{fig:CMS_pTnunu_unfolded}. This is important in that it supports searches using Monte Carlo simulation to model this important background.

\subsection{Searches in single-lepton $t\bar{t}$ final states}

The next result is a re-analysis of the single lepton $t\bar{t}$ + \ptmiss{} channel by ATLAS, targeting both stop squark pair production and simplified $t\bar{t}$+DM models \cite{ATLAS:2024rcx}. Compared with the previous published result, this included numerous improvements, including a new resolved top tagger, and dedicated neural networks targeting each signal. These neural networks were trained in a number of categories, and each category was then split into three based on NN score, with the lowest scores used as CRs of the major backgrounds (W boson, single top quark and $t\bar{t}$ production - the last in both single lepton and dileptonic decays), the intermediate scores as VRs, and the highest scores as SRs.

\begin{figure}[!ht]
    \centering
    \includegraphics[width=8cm]{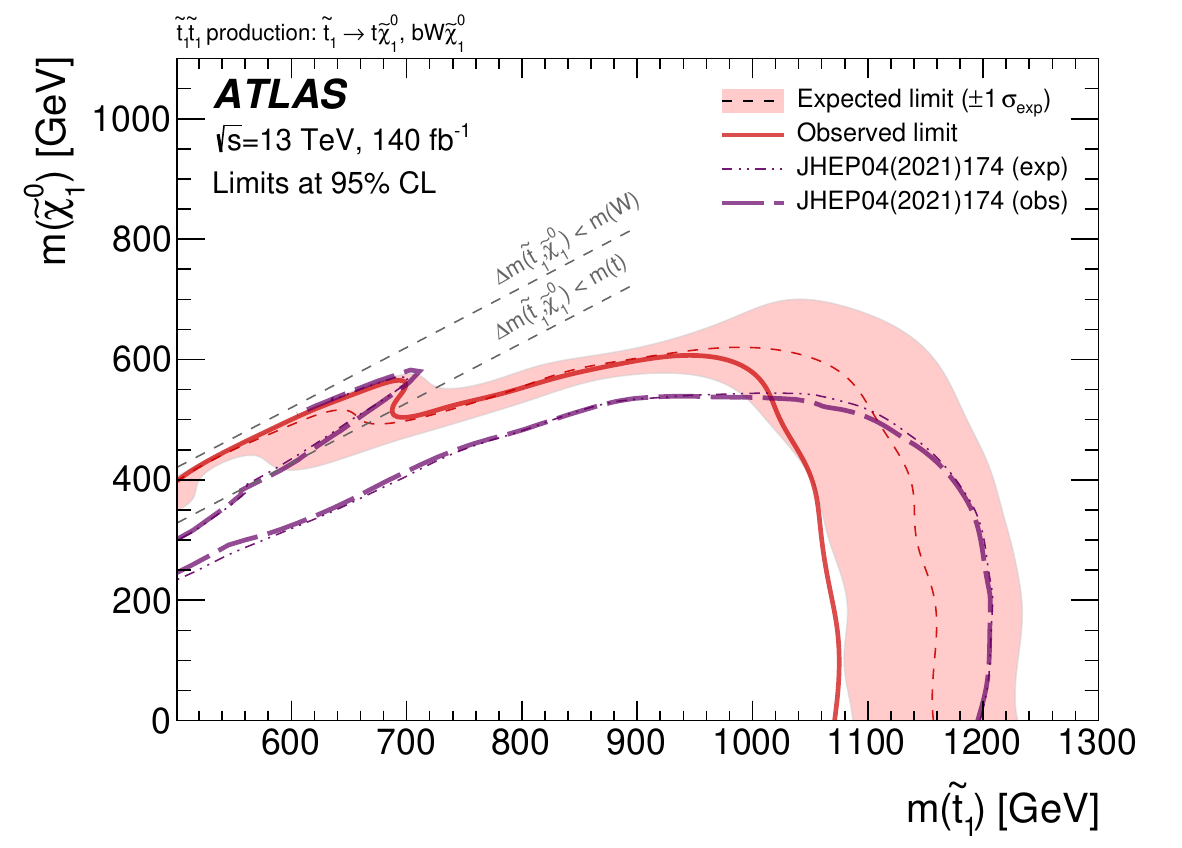}
    \caption{Limits on stop quark pair production from \cite{ATLAS:2024rcx} (red lines) compared with the previous result published in \cite{ATLAS:2020xzu} (purple lines).}
    \label{fig:ATLAS_SUSY_lims}
\end{figure}

No significant excess was observed (as was expected from the results of the previous analysis in this channel), and limits were set on both models. The limits for stop squark pair production, compared to the previous ATLAS search in this channel with the same dataset, are shown in figure \ref{fig:ATLAS_SUSY_lims}. The previous search used a large number of categories, which targeted both the compressed and boosted regions, and hence the new result only achieves the same level of sensitivity for the compressed region and slightly less for very boosted topologies. However the more general strategy, and the combination of resolved and boosted top tags, provided a significant improvement in sensitivity for the intermediate mass-gap scenario, where the old category-based analysis was particularly insensitive.

\begin{figure}[htbp]
\begin{minipage}[b]{.48\textwidth}
\begin{center}
\includegraphics[width=6cm]{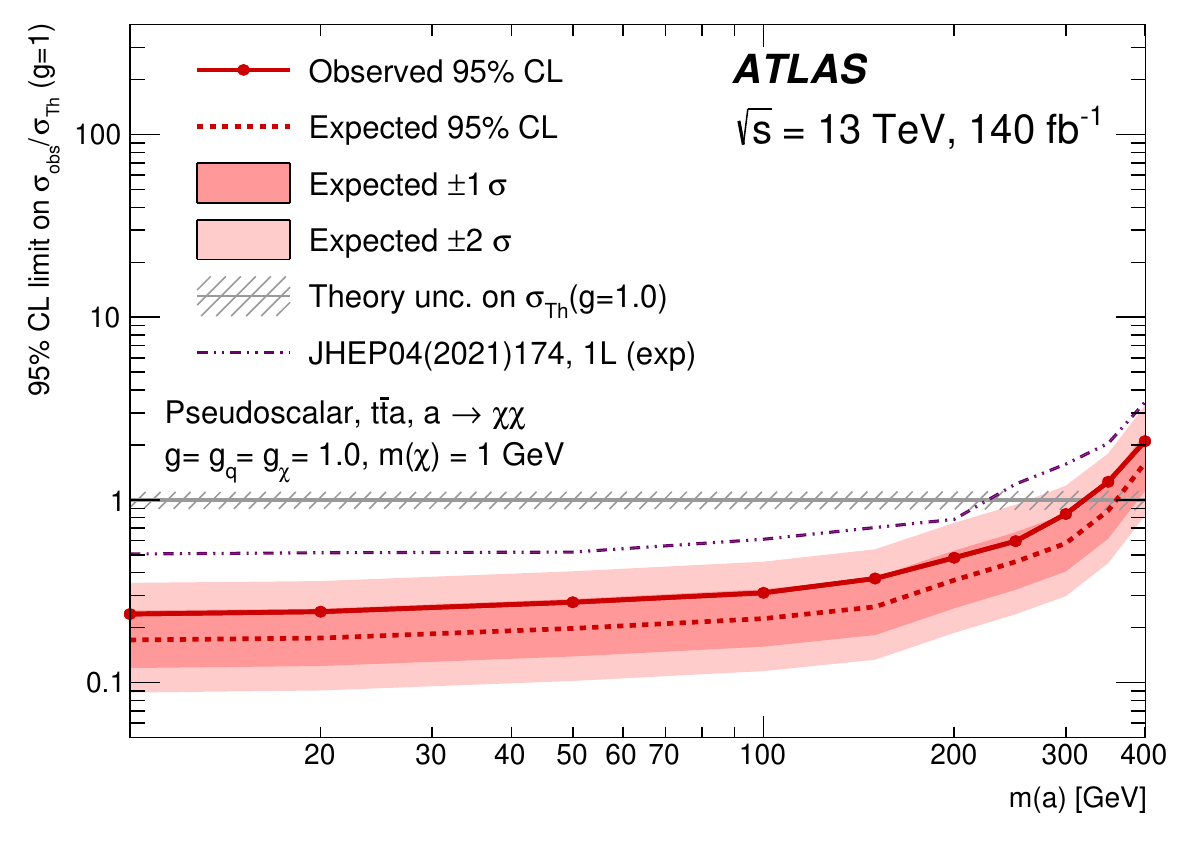}
\end{center}
\end{minipage} \hfill
\begin{minipage}[b]{.48\textwidth}
\begin{center}
\includegraphics[width=6cm]{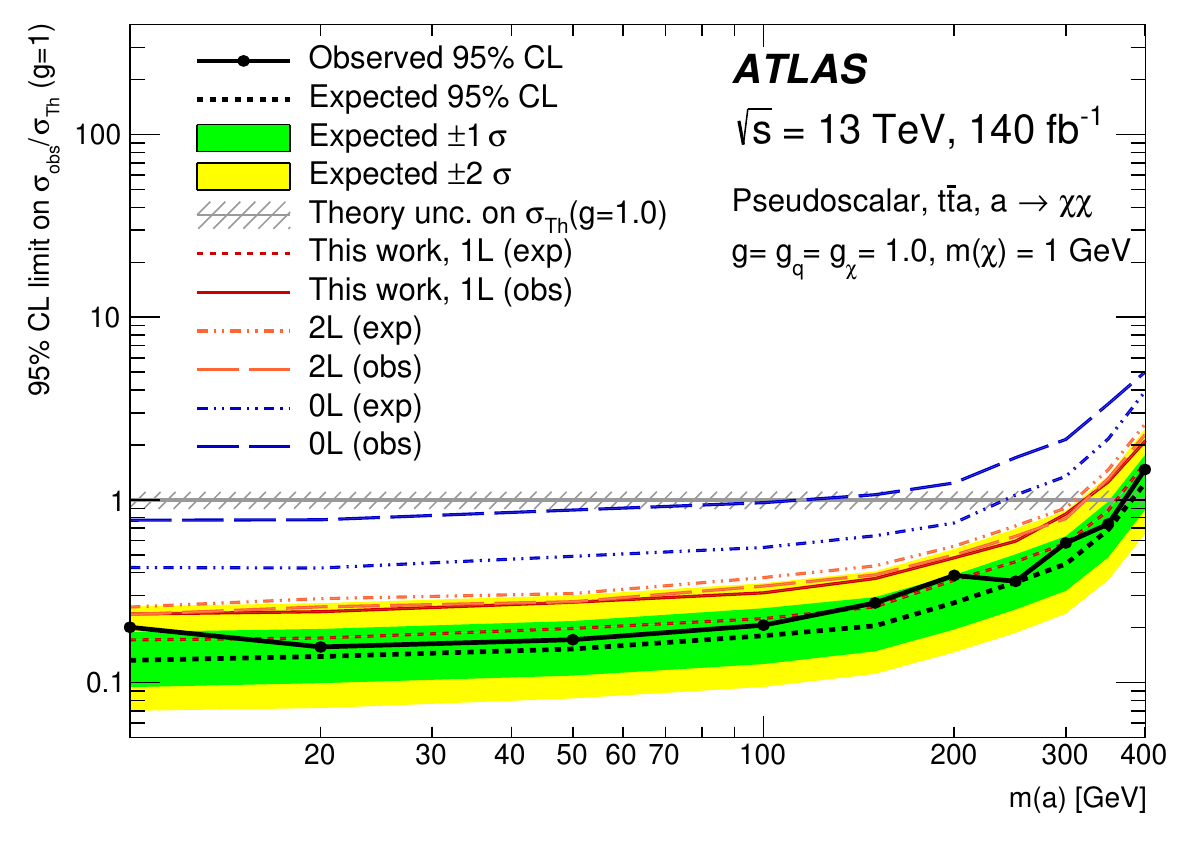}
\end{center}
\end{minipage} 
\caption{Limits on $t\bar{t}$+DM production via pseudoscalar mediator, in comparison with the previous results from \cite{ATLAS:2020xzu} (left) and combined with existing results in the 0 and 2 lepton channels (right).}
\label{fig:ATLAS_ttDM_lims}
\end{figure}

The results for the simplified $t\bar{t}$+DM model are shown in figure \ref{fig:ATLAS_ttDM_lims}; here the analysis shows a significant improvement compared to the previous result due to the dedicated neural network and resolved top taggers. This result was also combined with existing results in the 0 and 2 lepton channels, dramatically improving the sensitivity of the combination as this channel overtakes the dilepton channel to become most sensitive. 

\begin{figure}[htbp]
\begin{minipage}[b]{.48\textwidth}
\begin{center}
\includegraphics[width=7cm]{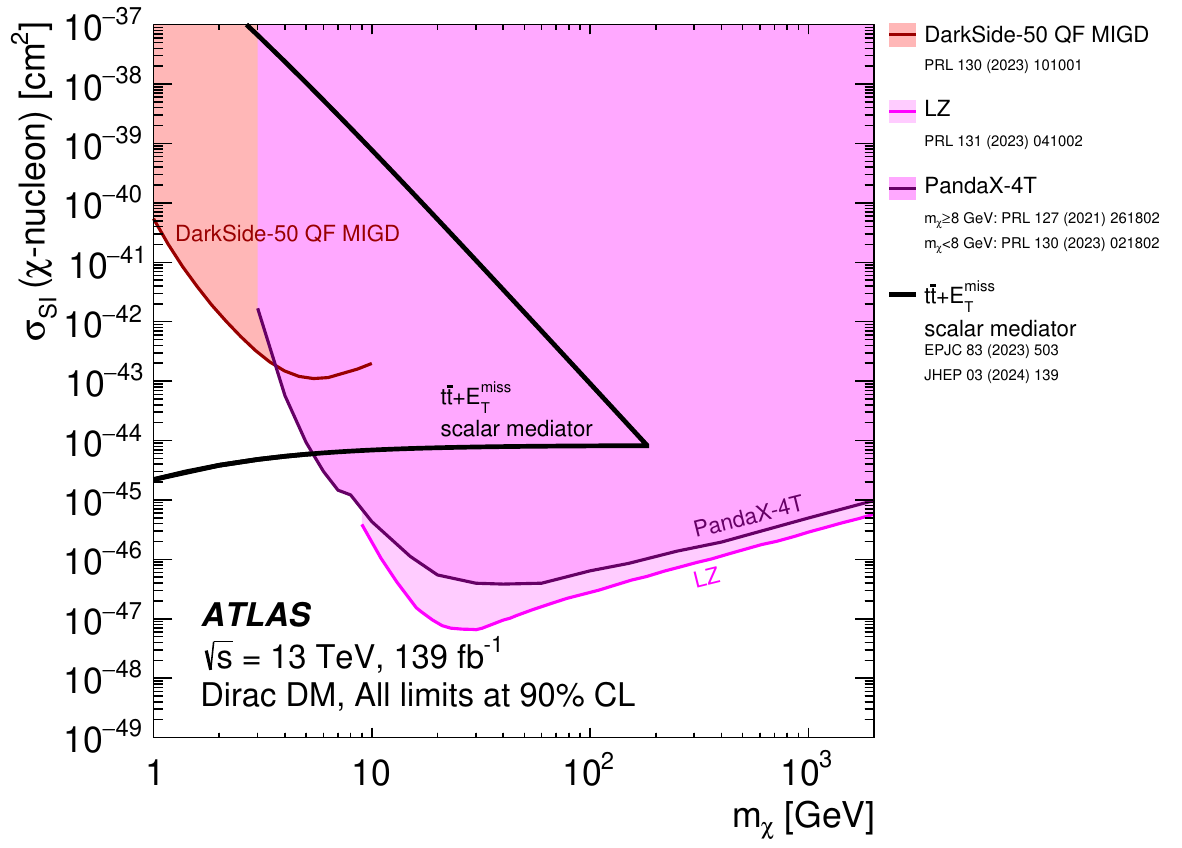}
\end{center}
\end{minipage} \hfill
\begin{minipage}[b]{.48\textwidth}
\begin{center}
\includegraphics[width=7cm]{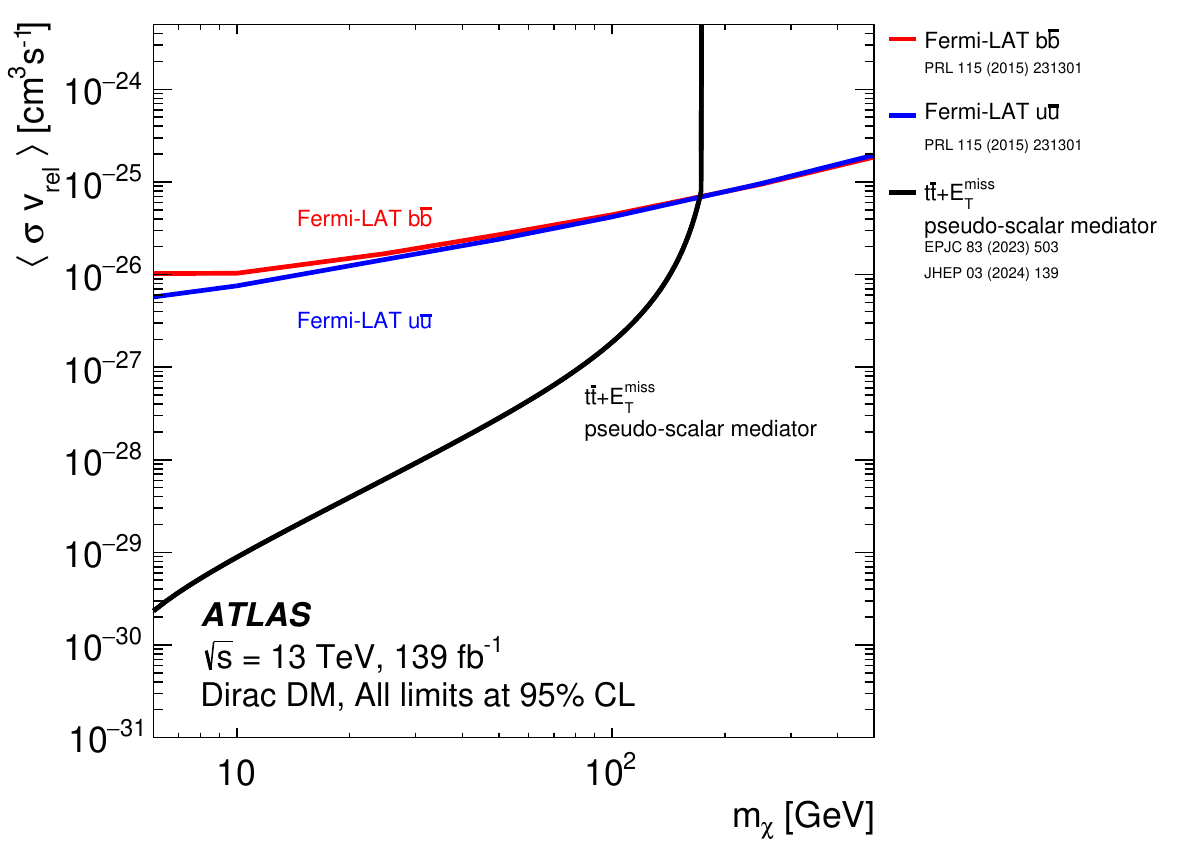}
\end{center}
\end{minipage} 
\caption{Limits on $t\bar{t}$+DM production, for the scalar mediator in comparison with leading direct detection experiments (left) and for the pseudoscalar mediator in comparison with leading indirect detection experiments (right).}
\label{fig:ATLAS_ttDM_comp_lims}
\end{figure}

Additionally, these combined results for these simplified $t\bar{t}$+DM models were compared with limits from other experiments, as shown in figure \ref{fig:ATLAS_ttDM_comp_lims}. For the scalar mediator, direct detection experiments, which search for DM in our galaxy's halo colliding with SM matter exciting emission of light (typically using tanks of Xenon well-shielded from external radiation), are the most sensitive for higher DM masses, but LHC experiments, which are sensitive to mediator mass but not the DM mass, are able to exclude the low DM mass region. For pseudoscalar mediators, direct detection experiments are much less sensitive due to the velocity-dependent nature of the interaction, so LHC limits are most sensitive within their mass reach, however for DM masses above half of the highest mediator mass covered, the best limits come from indirect detect experiments, which search for radiation coming from annihilation of DM in regions expected to have high density thereof, such as the galactic centre. 

\subsection{A charming Alternative: Stop squarks decaying to top and charm quarks}

\begin{figure}[htbp]
\begin{minipage}[b]{.48\textwidth}
\begin{center}
\includegraphics[height=4.5cm]{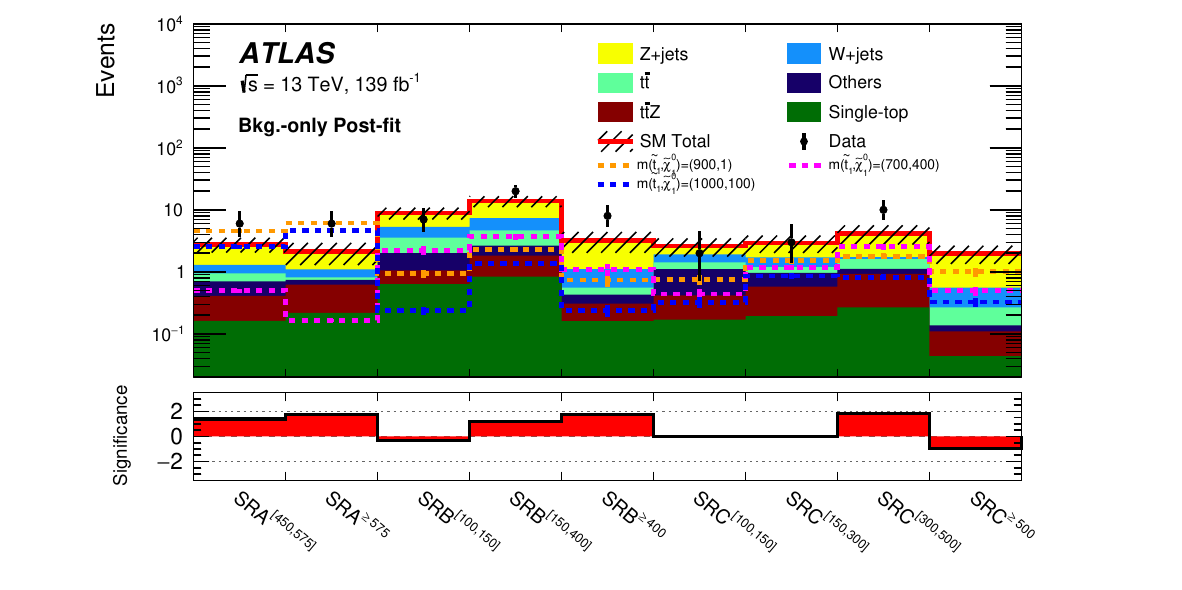}
\end{center}
\end{minipage} \hfill
\begin{minipage}[b]{.48\textwidth}
\begin{center}
\includegraphics[height=4.5cm]{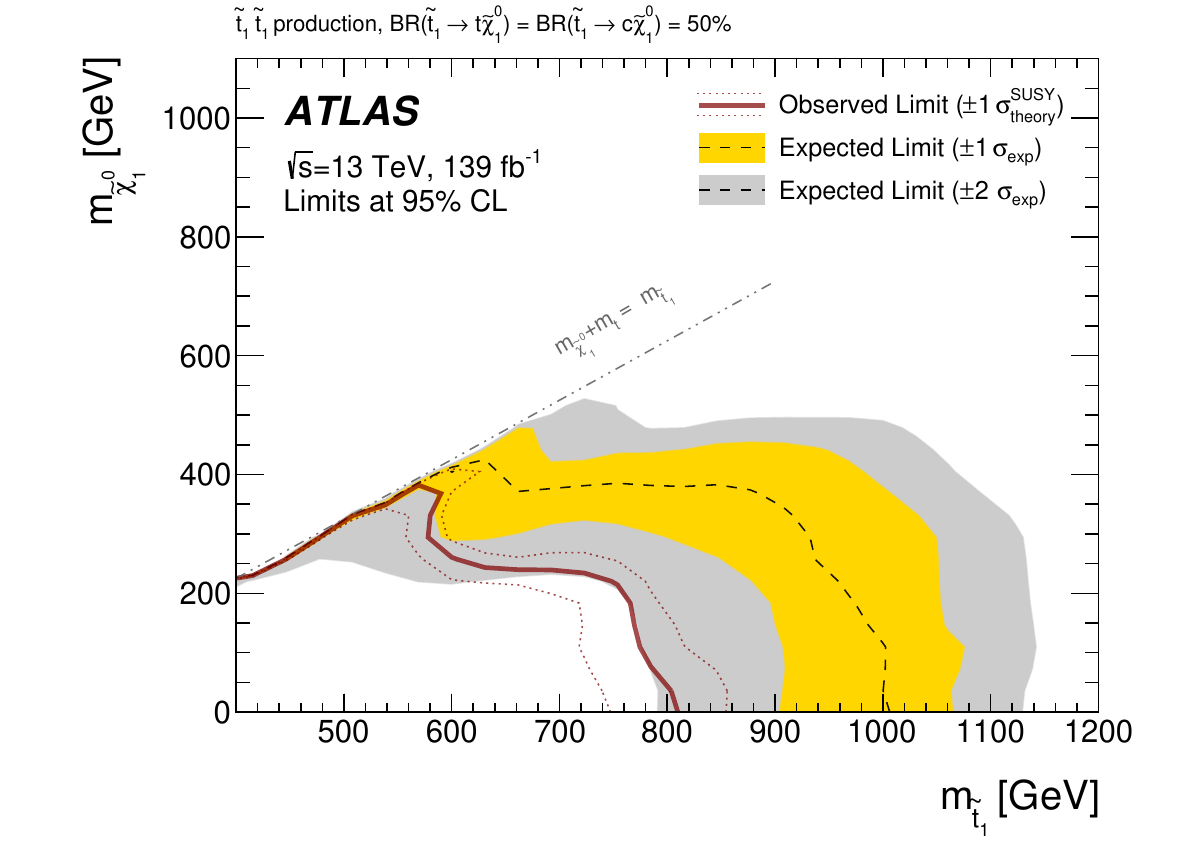}
\end{center}
\end{minipage} 
\caption{Left: Post-fit agreement in the SRs for flavour violating stop quark production in the boosted or intermediate phase space. Right: limits on this model.}
\label{fig:ATLAS_top_charm}
\end{figure}

In more exotic supersymmetric models, stop quarks may decay to other flavours of quark in addition to the top, such as the charm quark. A new search \cite{ATLAS:2024vyj} targets the case in which one stop squark decays to a neutralino and a (hadronically decaying) top quark, and the other to a charm quark. This uses boosted and resolved top taggers similar to \cite{ATLAS:2024rcx}, and a dedicated tagger for jets from charm quarks. This analysis used cuts on numerous kinematic variables to suppress common backgrounds, and then performed a fit in numerous signal categories targeting the boosted, intermediate and compressed topologies. Control regions were used to estimate the rates of major backgrounds, which for this analysis were single top, Z and W boson and $t\bar{t}$ production. The distribution in the boosted and intermediate SRs, and limits for this analysis are shown in figure \ref{fig:ATLAS_top_charm} - there is an excess in the boosted signal regions (SRA and SRB), which leads to an excess of with a significance of 2 $\sigma$ in the boosted part of the model phase space.

\subsection{Putting it all together: Dark matter produced in association with a single top quark or top quark pair}

\begin{figure}[htbp]
\begin{minipage}[b]{.32\textwidth}
\begin{center}
\includegraphics[width=4cm]{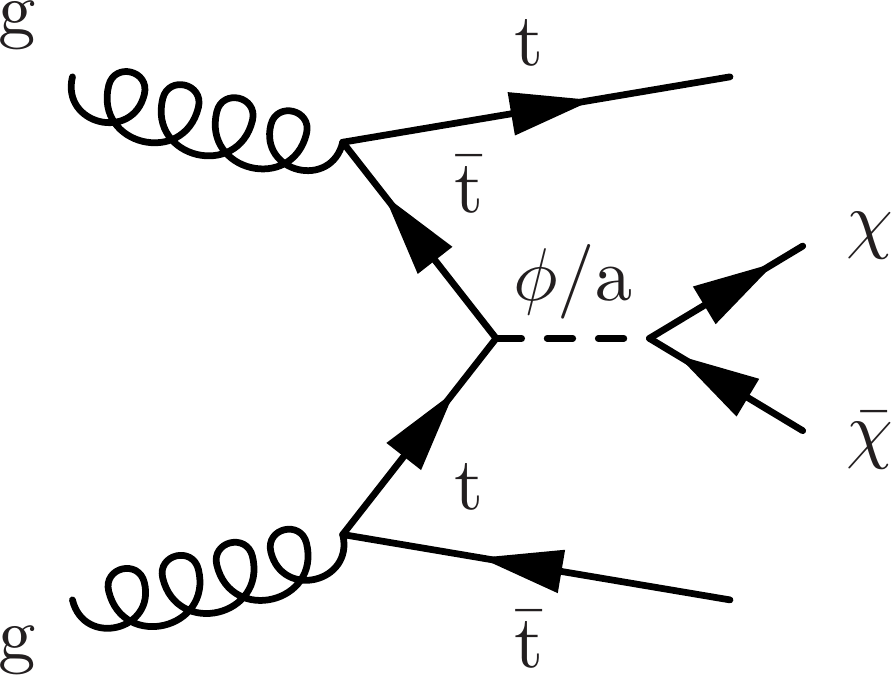}
\end{center}
\end{minipage} \hfill
\begin{minipage}[b]{.32\textwidth}
\begin{center}
\includegraphics[width=4cm]{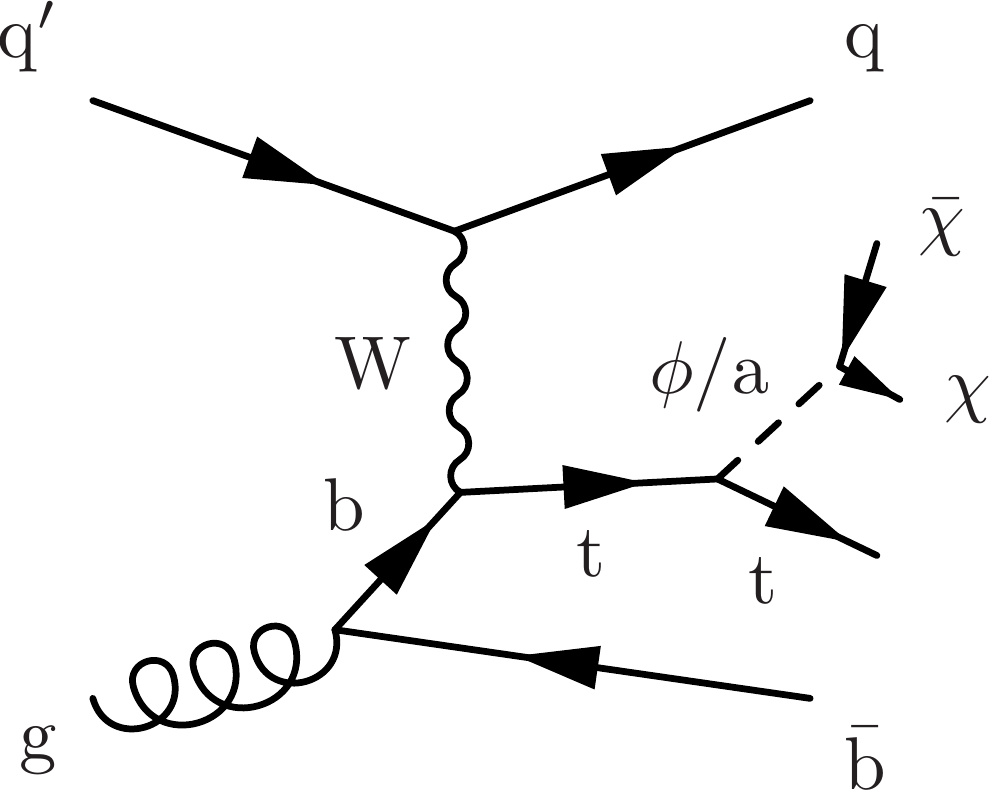}
\end{center}
\end{minipage} \hfill
\begin{minipage}[b]{.32\textwidth}
\begin{center}
\includegraphics[width=4cm]{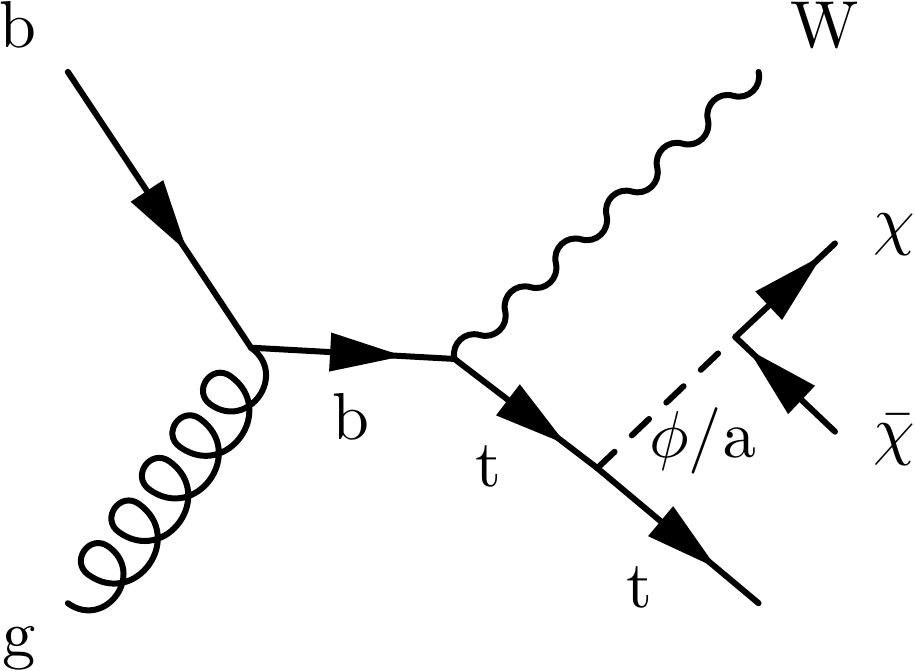}
\end{center}
\end{minipage}
\caption{Diagrams for production of DM in association with a top quark pair or a single top quark: left, $t\bar{t}$+DM production; centre, t-channel t+DM production, and right, tW+DM production.}
\label{fig:tttDM_diags}
\end{figure}

\begin{figure}[!ht]
    \centering
    \includegraphics[width=6cm]{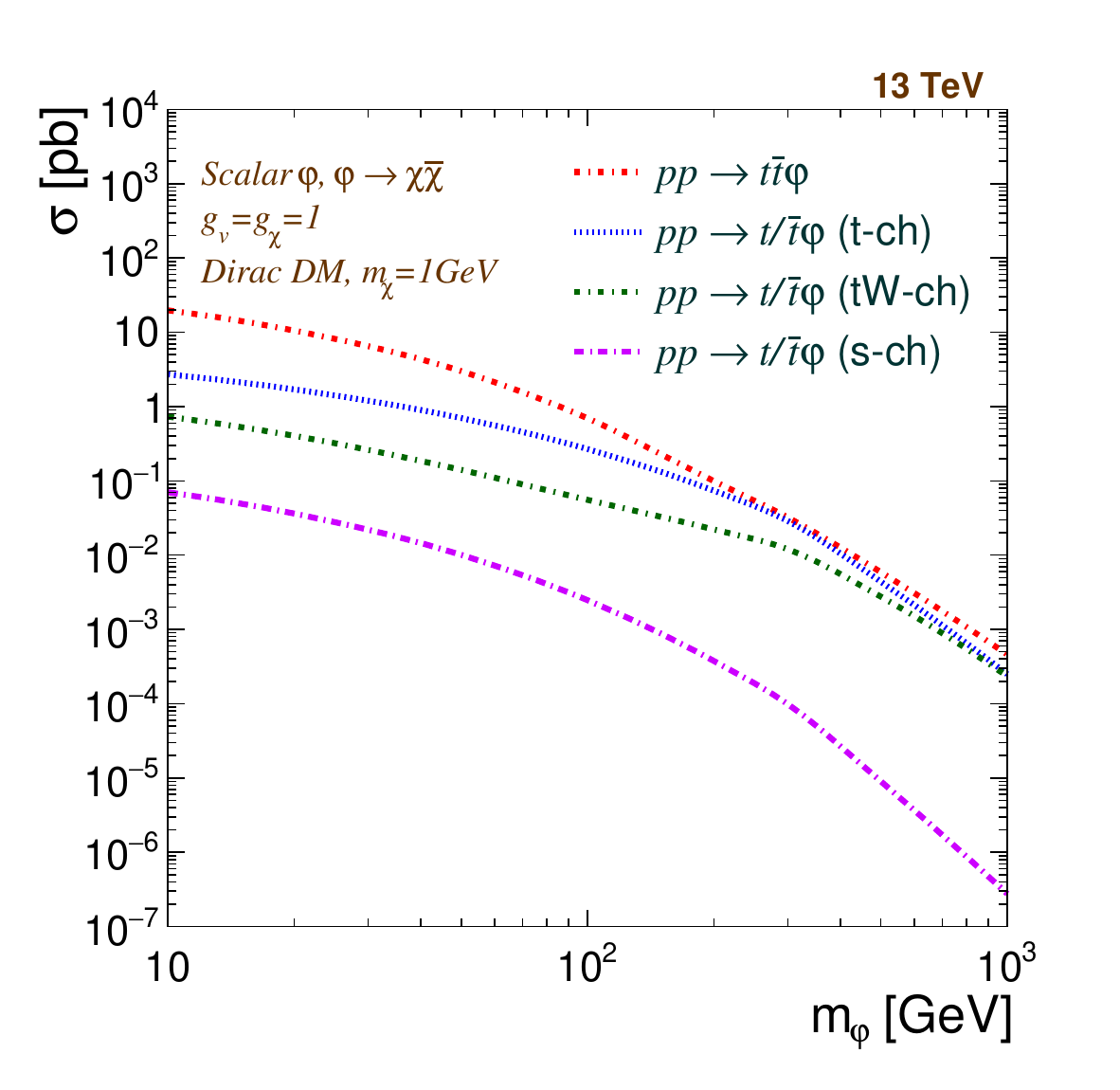}
    \caption{The cross section of different tt+DM and t+DM processes as a function of mediator mass for the scalar mediator.}
    \label{fig:tttDM_xs_comp}
\end{figure}

The simplified models which predict $t\bar{t}$+DM signatures also give rise to single top + DM (t+DM) diagrams, as shown in figure \ref{fig:tttDM_diags}. The cross section for these processes is lower than for $t\bar{t}$+DM production, but they decrease more slowly as a function of mediator mass since less energy is required to produce a single top than a $t\bar{t}$ pair, and for higher mediator masses the cross sections of both $tW$+DM and t-channel single top + DM production are comparable with $t\bar{t}$+DM, as can be seen in figure \ref{fig:tttDM_xs_comp}. Therefore CMS has recently released an analysis \cite{CMS:2024ybt} targeting both t+DM and $t\bar{t}$+DM across all top quark decay modes. It is worth noting that these t+DM topologies are significantly less boosted than those discussed in section \ref{sec:monotop}, and are more similar to $t\bar{t}$+DM with fewer jets and b jets. Accordingly this analysis requires fewer jets than analyses only targeting $t\bar{t}$+DM, and the number of b-tagged jets is used to categorise SRs targeting t+DM (1 b jet) and $t\bar{t}$+DM (2 b jets). Additionally in the 0 and 1 lepton channels the t+DM category is split into categories with zero and at least one forward jets to target t-channel single top + DM, which includes a spectator quark which often forms a jet in the forward part of the detector.

\begin{figure}[htbp]
\begin{minipage}[b]{.32\textwidth}
\begin{center}
\includegraphics[width=5cm]{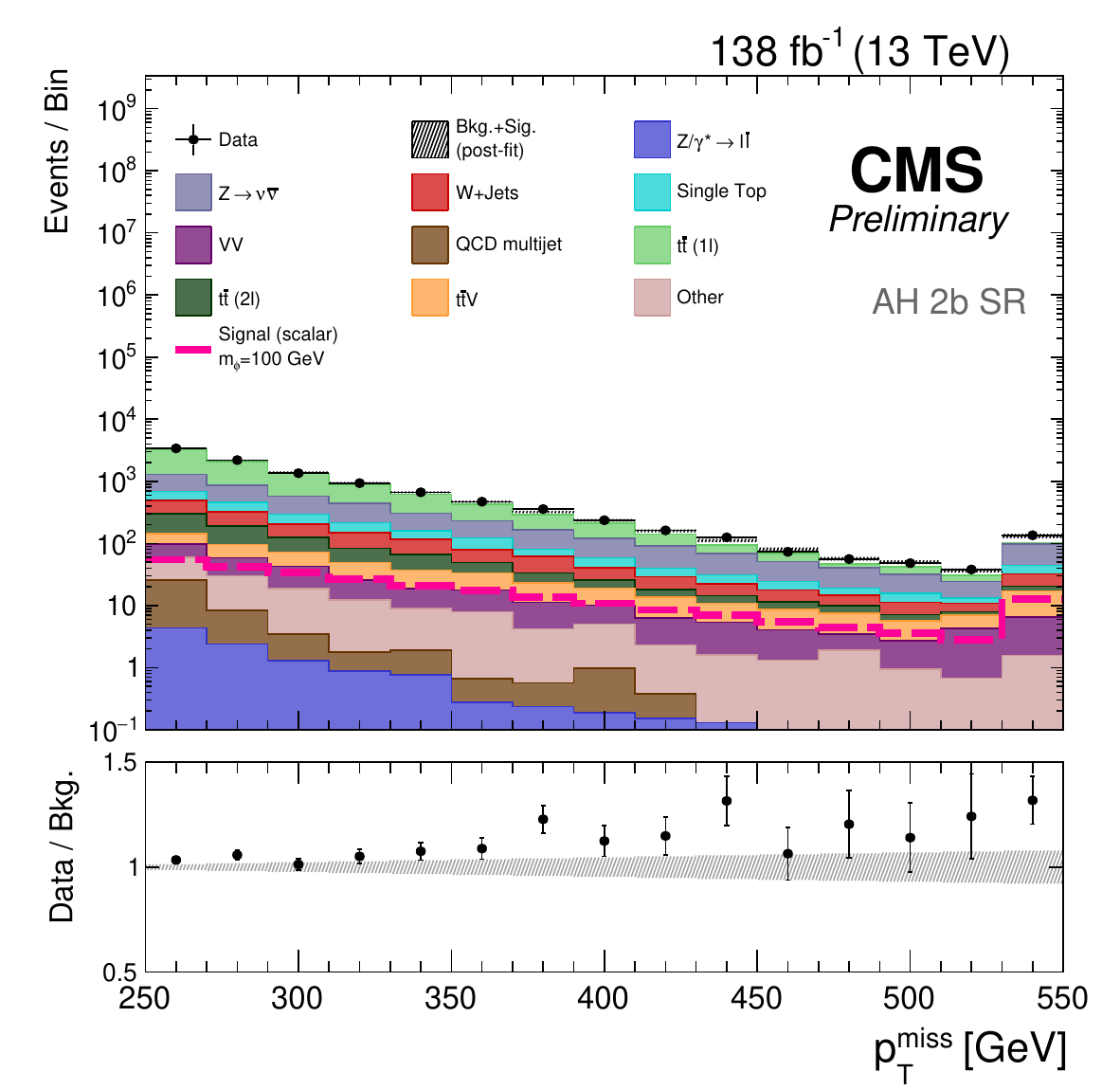}
\end{center}
\end{minipage} \hfill
\begin{minipage}[b]{.32\textwidth}
\begin{center}
\includegraphics[width=5cm]{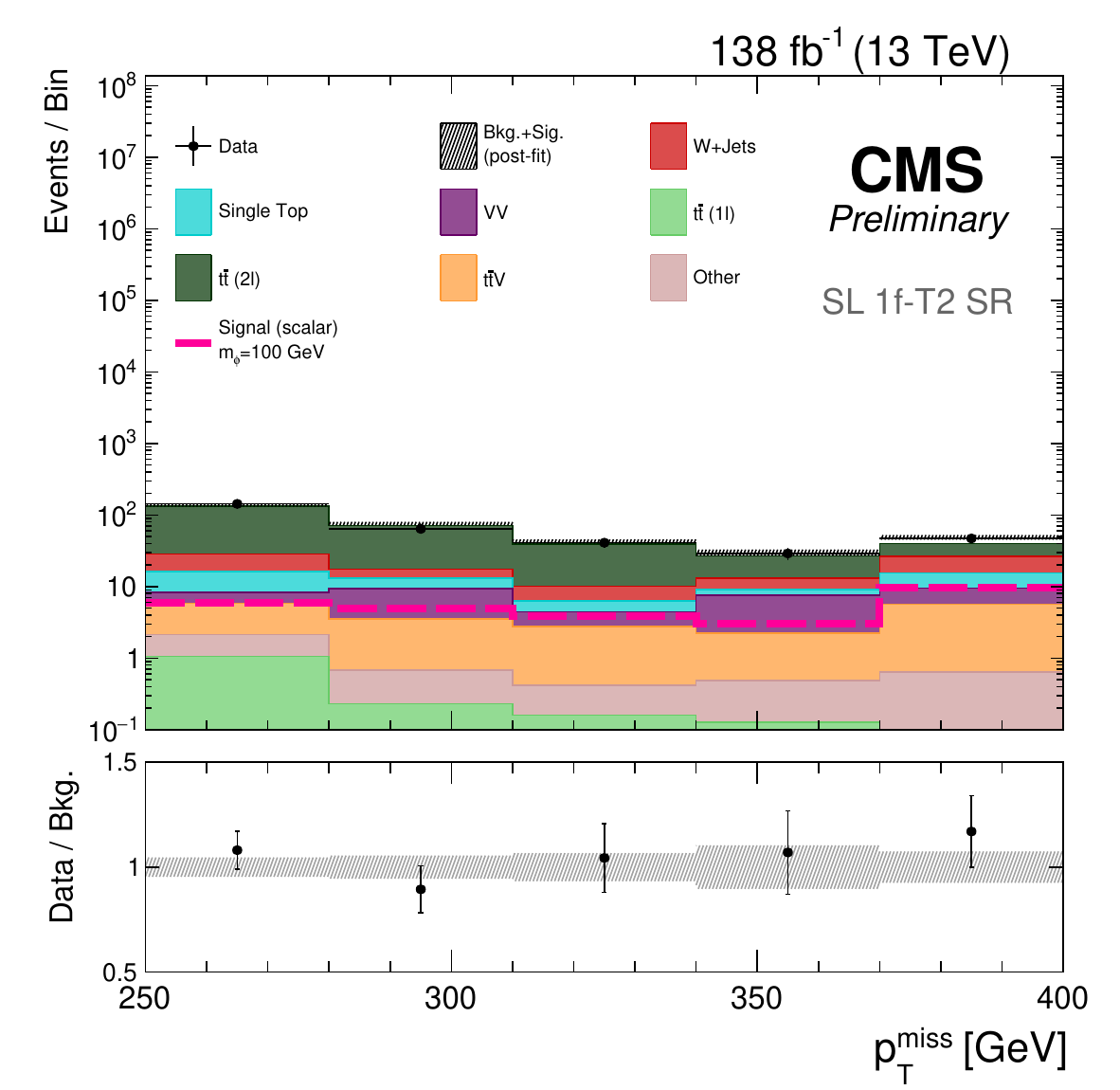}
\end{center}
\end{minipage} \hfill
\begin{minipage}[b]{.32\textwidth}
\begin{center}
\includegraphics[width=5cm]{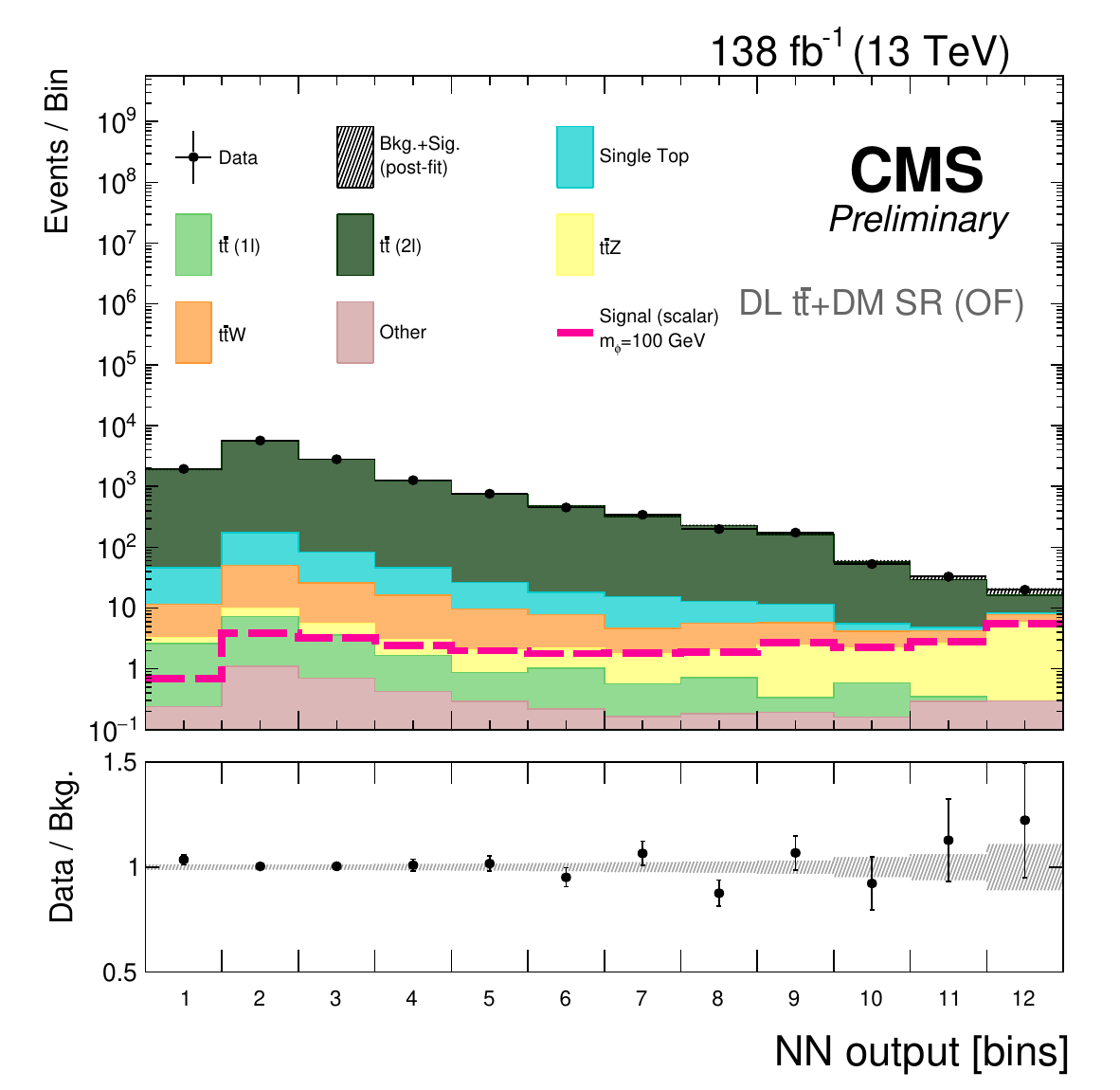}
\end{center}
\end{minipage}
\caption{Example SRs for the combined t+DM and $t\bar{t}$+DM search. Left, the 0 lepton, $\geq2$ b jet SR; centre, the 1 lepton, 1 b jet, $\geq1$ forward jet, high topness SR; right, the 2 lepton different flavour, $\geq2$ b jet SR.}
\label{fig:tttDM_SRs}
\end{figure}

\begin{figure}[!ht]
    \centering
    \includegraphics[width=8cm]{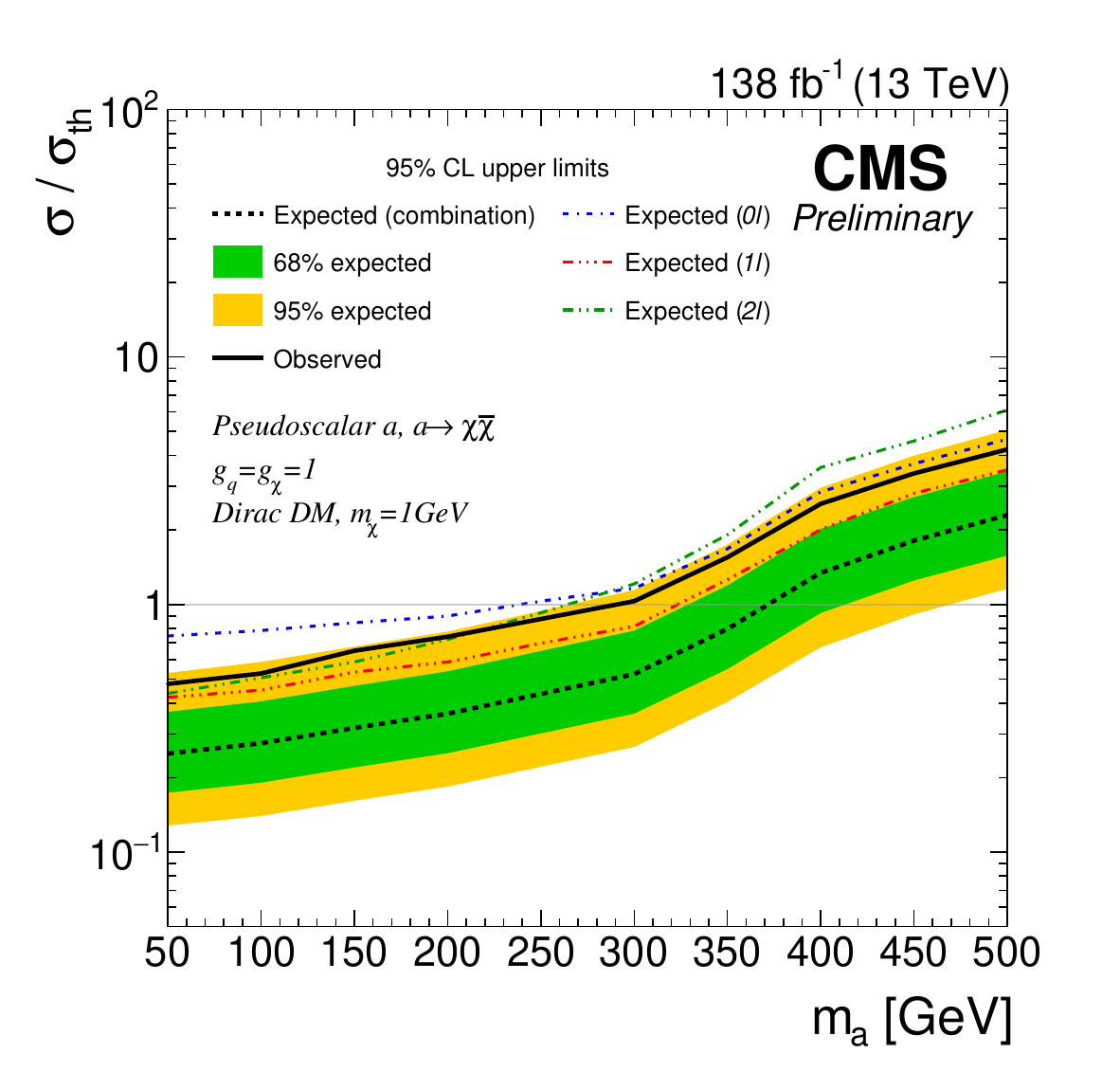}
    \caption{Limits on the combined cross section of t+DM and $t\bar{t}$+DM as a function of pseudoscalar mediator mass.}
    \label{fig:tttDM_lims}
\end{figure}

The different lepton channels face different challenges, and hence use different analysis strategies. The 0 and 1 lepton channels generally have high statistics but challenging backgrounds to model, and therefore cut on a series of kinematic variable and fit on \ptmiss{}, with numerous control regions linked to the SR in the final fit on a per-bin basis, allowing fine-grained estimation of the backgrounds. A categorisation on a variable called "topness", designed to suppress contributions from dileptonic $t\bar{t}$ where one lepton is not reconstructed, was used in the 1 lepton channel. \ptmiss{} is less sensitive in the 2 lepton channel, since the main background is dileptonic SM $t\bar{t}$ events containing two neutrinos, and so a simpler selection was used and the final fit was performed on the output of a NN. Some example SR distributions are shown in figure \ref{fig:tttDM_SRs}, and the limits on pseudoscalar mediator masses are shown in figure \ref{fig:tttDM_lims} - the most sensitive channel is 1 lepton, but the 2 lepton channel contributes significantly at low masses, while the 0 lepton channel makes a significant contribution at high masses. An excess of approximately 2 $\sigma$ significance is observed (with the largest excess in the 0 lepton channel). Since all signals peak at high \ptmiss{} (or NN score) this excess is consistent with all mediator masses considered, but is highest for mediators around 200 GeV.

\section{Conclusion}

Many models predict that dark matter could be produced in association with top quarks at the LHC. Several results have been published in the past year which search for these models and have been presented here. These analyses aim to maximise the potential of the Run 2 LHC dataset, using machine learning to improve signal extraction, and advanced data-driven background estimation techniques to confront the challenging phase space.

Despite some small excesses in certain analyses, there is not yet any clear evidence for DM production at the LHC. However there remains a significant amount of phase space to be explored in many models. Furthermore, many of the analyses presented here are still mainly limited by the limited statistics of the data, and the cross sections of high mediator mass processes increase notably even for the relatively small energy increases to 13.6 TeV for Run 3 of the LHC, and potentially 14 TeV in future runs. Therefore the prospects for future discoveries remain tantalising for Run 3 and beyond.

\section*{Acknowledgements}
The author would like to thank the ATLAS and CMS collaborations for the opportunity to present these interesting results, as well as the organisers of the Top Workshop for a well-run conference with an excellent scientific programme.

\paragraph{Funding information}
The author acknowledges support from DESY (Hamburg, Germany), a member of the Helmholtz Association HGF.

\bibliography{top+DM.bib}

\end{document}